\documentclass[acmsmall]{acmart}

\usepackage[ruled,linesnumbered,vlined]{algorithm2e}
\usepackage{amsmath}
\usepackage{amsfonts}
\usepackage{algpseudocode}
\usepackage{pifont}
\usepackage{color}
\usepackage{xcolor}
\usepackage{soul}
\usepackage{tabularx}
\usepackage{threeparttable}
\usepackage{makecell}
\usepackage{hyperref}
\usepackage{multirow}
\usepackage{nicematrix}

\AtBeginDocument{%
  \providecommand\BibTeX{{%
    \normalfont B\kern-0.5em{\scshape i\kern-0.25em b}\kern-0.8em\TeX}}}

\setcopyright{acmcopyright}
\copyrightyear{2023}
\acmYear{2023}
\acmDOI{https://doi.org/10.1145/3568992}

\renewcommand\footnotetextcopyrightpermission[1]{}
\acmJournal{TRETS}
\acmVolume{16}
\acmNumber{3}

\acmSubmissionID{123-A56-BU3}

\begin{document}

\title{fSEAD: a Composable FPGA-based Streaming Ensemble Anomaly Detection Library}

\author{Binglei Lou}
\affiliation{%
  \institution{The University of Sydney}
  \streetaddress{Camperdown NSW 2006}
  \city{Sydney}
  \country{Australia}}
\email{binglei.lou@sydney.edu.au}

\author{David Boland}
\affiliation{%
  \institution{The University of Sydney}
  \streetaddress{Camperdown NSW 2006}
  \city{Sydney}
  \country{Australia}}
\email{david.boland@sydney.edu.au}

\author{Philip H.W. Leong}
\affiliation{%
 \institution{The University of Sydney}
 \streetaddress{Camperdown NSW 2006}
 \city{Sydney}
 \country{Australia}}
\email{philip.leong@sydney.edu.au}


\begin{abstract}
  Machine learning ensembles combine multiple base models to produce a more accurate output. They can be applied to a range of machine learning problems, including anomaly detection. In this paper, we investigate how to maximize the composability and scalability of an FPGA-based streaming ensemble anomaly detector (fSEAD). To achieve this, we propose a flexible computing architecture consisting of multiple partially reconfigurable regions, pblocks, which each implement anomaly detectors. Our proof-of-concept design supports three state-of-the-art anomaly detection algorithms: Loda, RS-Hash and xStream. Each algorithm is scalable, meaning multiple instances can be placed within a pblock to improve performance. Moreover, fSEAD is implemented using High-level synthesis (HLS), meaning further custom anomaly detectors can be supported. Pblocks are interconnected via an AXI-switch, enabling them to be composed in an arbitrary fashion before combining and merging results at run-time to create an ensemble that maximizes the use of FPGA resources and accuracy. Through utilizing reconfigurable Dynamic Function eXchange (DFX), the detector can be modified at run-time to adapt to changing environmental conditions. We compare fSEAD to an equivalent central processing unit (CPU) implementation using four standard datasets, with speed-ups ranging from $3\times$ to $8\times$.

\end{abstract}

\ccsdesc[500]{Hardware~Reconfigurable logic applications}
\ccsdesc[500]{Computing methodologies~Ensemble methods}
\ccsdesc[300]{Hardware~Hardware accelerators}

\keywords{FPGA, Anomaly Detection, Partial Reconfiguration, Composability}

\maketitle

\section{Introduction}
\label{se:introduction}
Anomaly detection is a key machine learning (ML) task and refers to the automatic identification of unforeseen or abnormal samples embedded in normal data~\cite{survey2009,kernal2007}. Applications of anomaly detection include fault detection surveillance systems~\cite{surveillance2014}, fraud detection in financial transactions~\cite{finance2016}, intrusion detection for network security~\cite{security2009}, monitoring of sensor readings in aircraft~\cite{aircraft2019} and discovery of potential risks or medical problems in health data with predictive maintenance~\cite{medical2014}. 

Ensembles are a class of methods that pool weak detectors to form a more accurate combination~\cite{Freund1996}. Over a number of decades, they have proven to be an excellent methodology which utilises diversity of weak detectors to reach a better overall decision than the individual ones~\cite{Freund1997}. Each sub-detector in an ensemble is data-independent with identical structure. This makes it naturally amendable for parallel processing to obtain high throughput. The ensemble-size, i.e. the number of sub-detectors, is determined according to the computational resources and constraints of system performance. Generally, executing a large ensemble on central processing units (CPUs) as the sequential nature of program execution is mismatched to the available parallelism. Field-Programmable Gate Array (FPGA) is an attractive solution since spatial parallelism can be employed. 

When data is susceptible to concept drift~\cite{concept_drift}, anomaly detectors can be utilised in a streaming fashion, with the detector being updated in an online manner. Streaming techniques store and process a window of recent instances. Streaming ensemble anomaly detectors (SEADs) achieve high accuracy under limited memory, processing and time constraints because ensembles contribute high accuracy and robustness, and real-time processing is enabled via streaming algorithms. 

Existing anomaly detection libraries have been developed for CPUs. These include unsupervised, supervised, heterogeneous approaches such as SUOD~\cite{suod2021} and PyOD~\cite{pyod2019}. These libraries provide a single, well-documented application programming interface (API), making it easy to compare and compose different algorithms. In particular, PySAD introduces a framework of streaming ADs in Python~\cite{pysad2020} with a common interface. Using the aforementioned AD libraries, a software based model combination toolkit: \(combo\) was presented in reference~\cite{combo2020}, allowing anomaly detectors to be combined in Python. 
CPU-based SEAD libraries allows programmers to switch detector types or to combine multiple detectors to boost performance for specific scenarios. 

Implementing customized anomaly detection algorithms in hardware is desirable to achieve higher performance, lower power and lower latency~\cite{pca_fpga2008,niclof2017,autoencoder_fpga2018}. However, most designs do not have comparable flexibility or customisability as software based approaches. This is one of the main challenges for reconfigurable computing and while we do not solve the problem, we propose an approach with considerably more flexibility than conventional FPGA designs. To the best of our knowledge, no composable FPGA implementations of ADs have been published to date.

In this paper, to enable AD on FPGA with higher flexibility and scalability, we propose fSEAD, a composable and low latency FPGA-based SEAD library. fSEAD has two components. The first is a high-level synthesis (HLS) based module generator which converts three state-of-the-art SEAD algorithms (Loda~\cite{loda2016}, RS-Hash~\cite{rshash2016}, and xStream~\cite{xstream2018}) into optimized sub-detector-level paralleled FPGA entities which store all parameters in on-chip memory. The second component is a composable hardware framework that enables online switching and dynamic routing of data between IP cores at run-time through reconfigurable Dynamic Function eXchange (DFX) and the arbitrary routable AXI4-Stream Switches. This allows new functionality to be introduced to the design at run-time. Composability is gained through the combination of coarse-grained reconfigurability of sub-detectors and switchability which facilitates their flexible interconnection. While run-time reconfiguration of the FPGA introduces tens or hundreds of milliseconds in overhead for large FPGAs~\cite{xilinx_ICFPT,QCD2022}, this is not a concern for fSEAD as this is only done when fSEAD is idle. The main contributions of this paper are:

\begin{itemize}

\item The first FPGA-based ML system that allows complex and more powerful ADs to be created from simple blocks without recompilation. 

\item The creation of hardware implementations for three streaming ensemble anomaly detectors (Loda, RS-Hash and xStream) and demonstration of how they can be integrated within our framework. These implementations are created from an HLS-based generator for FPGA instances, with a GCC-based alternative that creates multi-threaded CPU versions for comparison. New detectors can be written in C and Python and are easily integrated in this library.

\item A composable framework which utilizes multiple reconfigurable regions connected via two AXI switches. In the implementation, high frequency is achieved through floorplanning of pblocks which surround the fSEAD infrastructure, minimising routing delay.

\item We use PYNQ partial overlays invoked in Python for executing AD functions~\cite{PYNQ}, allowing an easy-to-use interface for composing and comparing different ADs.

\item This research uses publicly available datasets and our design has been made open source to facilitate reproducible research~\footnote{fSEAD: \url{https://github.com/bingleilou/fSEAD}.}.
\end{itemize}

The remainder of the article is organised as follows. We start in Section~\ref{se:2_background} with an introduction of the SEAD background and a formal definition of the framework that we deployed in the FPGA. In Section~\ref{se:3_fsead_methodology}, we describe the design of the proposed fSEAD from four aspects: Module Generator, DFX Tool Flow, Composable Infrastructure and its FPGA Implementation. Then, the results of experiment and performance are showed in Section~\ref{se:4_experiment}. Finally, Section~\ref{se:5_conclusion} conclude the paper and discuss future work.

\section{Background and Related Work}
\label{se:2_background}
In this section, we first introduce anomaly detectors for streaming data. We then discuss how they can be combined using ensemble-centric methods to achieve greater accuracy. We highlight how this has led to the development of several comprehensive software-based AD libraries; this illustrates the desire for our flexible hardware accelerated library which can achieve better performance and similar accuracy to these software implementations. Finally, we discuss the Dynamic Function eXchange (DFX) technique and partial overlays, which we have used to provide the flexibility to choose different ensembles at run-time in our accelerator.

\subsection{Streaming Anomaly Detection}
\label{se:2_sead}
Anomaly detection is a key machine learning (ML) task, which refers to the automatic identification of unforeseen or abnormal samples embedded in a large amount of normal data~\cite{survey2009,kernal2007}. From the perspective of processing data, we distinguish between two anomaly detection types: static and streaming. 

Static detectors usually operate on a relatively large batch of data before performing information extraction and feature analysis to identify rare items, events or observations from the general distribution of a population.  Representative methods include k-Nearest Neighbors (kNN)~\cite{KNN2000}, Local Outlier Factor (LOF)~\cite{LOF2000}, and Principal Component Analysis (PCA)~\cite{PCA_ad2003}; for a more comprehensive collection of static methods, we refer the reader to the SUOD~\cite{suod2021}. While batch processing can lead to high throughput and accuracy, it is not suitable for systems that require real-time performance, since the computing resource requirements and latency increase with batch size.

In contrast, streaming methods only store and process a window of recent instances ~\cite{kernal2007,pysad2020}.  This is more amenable to achieving accurate anomaly scores under limited memory, processing and time constraints. Moreover, the algorithms are designed to facilitate more light-weight and potentially real-time implementations. A group of algorithms that support streaming anomaly detection processing includes, but is not limited to, ensemble-centric methods~\cite{loda2016,rshash2016,xstream2018}, tree-based methods~\cite{iForest2008,RSForest2014,HSTree2011}, kernel-methods~\cite{kernal2007}, as well as Neural Network-based solutions such auto-encoders~\cite{autoencoder_ad2014} and adversarial models~\cite{GAN_ad2019}. 

\begin{table}
\caption{Block Diagram of SEAD Methods}
\begin{threeparttable}[t]
\begin{tabular}{|c|c|c|c|c|}
\hline
\textbf{Anomaly}&\multicolumn{4}{c|}{\textbf{Blocks}} \\
\cline{2-5} 
\textbf{Detectors} & \textbf{\textit{Projection}}& \textbf{\textit{Core}}& \textbf{\textit{Sliding-Window}}& \textbf{\textit{Score}} \\
\hline
Loda & prj-loda & histogram & $1 \times W$ & $- {\log _2}(c/W)$ \\
\hline
RS-Hash & prj-rshash & CMS & $w \times W$ & $- {\log _2}(1 + \min \left\{ {{c_1}, \ldots ,{c_W}} \right\})$ \\
\hline
xStream & prj-xstream & CMS & $w \times W$ & $- {\log _2}(1 + \min \left\{ {{2^1}{c_1}, \ldots ,{2^W}{c_W}} \right\})$ \\
\hline

\end{tabular}

\begin{tablenotes}
    \footnotesize
    \item[1] \(W\): the length of the sliding-window.
    \item[2] \(w\): the number of hash functions in the count-min sketch (CMS).
    \item[3] \(c\): the count of histogram or hash code of CMS
\end{tablenotes}

\end{threeparttable}
\label{tb:sead_methods}
\end{table}

In this article, we select a set of three states of the art and representative anomaly detectors for our HLS module generator: Loda (Light-weight Online Detector of Anomalies)~\cite{loda2016}, RS-Hash~\cite{rshash2016}, and xStream (Outlier Detection in Feature-Evolving Data Streams)~\cite{xstream2018}. These are briefly explained below. We denote $X = \{ \vec{x}_i \}$ as the input dataset and dimension \(d\) of each sample, then $\vec{x}_i \in \mathbb{R}^d$ is the \(ith\) input sample of the data stream, and $R$ is the ensemble size.

Loda is a projection-based histogram detector, RS-Hash uses a randomized subspace hashing algorithm, while xStream is a density-based detector. Although the three techniques are based on different principles, the algorithms can be expressed as a composition of the following standardized blocks: \textit{Projection}, \textit{Core}, \textit{Sliding-window}, and \textit{Score}. Table~\ref{tb:sead_methods} summarizes the main functional blocks of these three SEAD methods. 

\begin{algorithm}
\SetAlgoLined
\DontPrintSemicolon
	\caption{Loda}
	\label{al:loda}
		\KwIn{Streaming input signal ${X} = \vec{x}_i \in \mathbb{R}^d$; ensemble size $R$; data dimension $d$; sliding-window length $W$.}
		\KwOut{Streaming anomaly value $Score$.}
		    \#pragma HLS DATAFLOW\\
		    // \ding{182}\textsc{Windower}:\\
		    \For {$dim=1,2,\ldots,d$}{
                A shift register to produce an entire sample: $X$ with \(d\) features.\\
            }
            // \ding{183}\textsc{Ensemble}:\\
    		\For {$r=1,2,\ldots,R$}{
    		    // \ding{184}\textsc{Projection}:\\
    			\For {$dim=1,2,\ldots,d$}{
    			    \#pragma HLS PIPELINE\\
    				$prj\_X \leftarrow prj\_X + X*loda\_prj$ (random projection)\\
    			}
                // \ding{185}\textsc{Histogram}:\\
    			$idx \leftarrow (prj\_X - loda\_\min )/(loda\_\max  - loda\_\min )$\\
    			$v \leftarrow sliding$-$window[idx]$\\
    			// \ding{186}\textsc{Sliding-window}:\\
    			Update $sliding$-$window$.\\
    			// \ding{187}\textsc{Score}:\\
    			${score(r)} \leftarrow {\log _2}\left( {v} \right)$\\
    		}
    		
    	// \ding{188}\textsc{Score Averaging}:\\
		$ Score \leftarrow \frac{1}{R}\sum\limits_{r = 1}^R {score(r)} $\\
		\Return{$Score$}
\end{algorithm}

\begin{algorithm}
\SetAlgoLined
\DontPrintSemicolon
	\caption{RS-Hash}
	\label{al:rshash}
		\KwIn{Streaming input signal ${X} = \vec{x}_i \in \mathbb{R}^d$; ensemble size $R$; data dimension $d$; sliding-window length $W$; hash functions number in CMS: $w$.}
		\KwOut{Streaming anomaly value $Score$.}
		    \#pragma HLS DATAFLOW\\
		    // \ding{182}\textsc{Windower}:\\
		    \For {$dim=1,2,\ldots,d$}{
                A shift register to produce an entire sample: $X$ with \(d\) features.\\
            }
            // \ding{183}\textsc{Ensemble}:\\
    		\For{$r=1,2,\ldots,R$}{
    		    // \ding{184}\textsc{Projection}:\\
    			\For{$dim=1,2,\ldots,d$}{
    			    \#pragma HLS PIPELINE\\
    			    $norm\_X \leftarrow$  normalize $X[\dim ]$ to the range of [0,1]\\
    				$prj\_X \leftarrow (norm\_X + rshash\_alpha[r][\dim ])/rshash\_f[r]$\\
    			}
    			// \ding{185}\textsc{Hash-Function}:\\ 
    			\For{$row=1,2,\ldots,w$}{
    			    \#pragma HLS UNROLL\\
    			    $prj\_hash[row][\dim ] \leftarrow prj\_X$\\
				    $hash\_value[row] \leftarrow Jenkins(key=prj\_hash[row],len=d,seed=row)$\\ 
				    (see \textbf{Algorithm~\ref{al:jenkins}} for details of $Jenkins$)\\
				    ${v_{row}} \leftarrow sliding$-$window[hash\_value[row]]$\\
				    // \ding{186}\textsc{Sliding-window}:\\
    			    Update $sliding$-$window$.\\
    			}
    			$min\_v \leftarrow \min \{ {v_1},{v_2},...,{v_{w}}\} $\\
    			\tcp{\ding{187}\textsc{Score}:}
                ${score(r)} \leftarrow {\log _2}\left( {min\_v} \right)$\\
    		}
    	// \ding{188}\textsc{Score Averaging}:\\
		$ Score \leftarrow \frac{1}{R}\sum\limits_{r = 1}^R {score(r)} $\\
		\Return{$Score$}
\end{algorithm}

\begin{algorithm}
\SetAlgoLined
\DontPrintSemicolon
	\caption{xStream}
	\label{al:xstream}
		\KwIn{Streaming input signal ${X} = \vec{x}_i \in \mathbb{R}^d$; ensemble size $R$; data dimension $d$; sliding-window length $W$; hash functions number in CMS: $w$; projection size: $K$.}
		\KwOut{Streaming anomaly value $Score$.}
		    \#pragma HLS DATAFLOW\\
		    // \ding{182}\textsc{Windower}:\\
		    \For{$dim=1,2,\ldots,d$} 
		    {
                A shift register to produce an entire sample: $X$ with \(d\) features.
            }
            // \ding{183}\textsc{Ensemble}:\\
    		\For{$r=1,2,\ldots,R$}{
    		    // \ding{184}\textsc{Projection}:\\
    			\For {$dim=1,2,\ldots,d$}{
    			\#pragma HLS PIPELINE\\
    				\For {$k=1,2,\ldots,K$}{
    				    \#pragma HLS UNROLL\\
    				     $prj\_X[k] \leftarrow prj\_X[k] + X[\dim ]*xstream\_prj[\dim ][k]$
    				}
    			}
				// \ding{185}\textsc{Hash-Function}:\\
    			\For {$row=1,2,\ldots,w$}{
    			    \#pragma HLS UNROLL\\
    			    $prj\_hash[row] \leftarrow perbins(prj\_X)$\\
				    $hash\_value[row] \leftarrow Jenkins(key=prj\_hash[row],len=K,seed=row)$ \\
				    (see Reference~\cite{xstream2018} and \textbf{Algorithm~\ref{al:jenkins}} for details of $perbins$ and $Jenkins$)\\
				    ${v_{row}} \leftarrow sliding$-$window[hash\_value[row]]$\\
				    // \ding{186}\textsc{Sliding-window}:\\
    			    Update $sliding$-$window$.\\
    			    $scor{e_{row}} \leftarrow {\log _2}({v_{row}}) + row$\\
    			}
    			// \ding{187}\textsc{Score}:\\
    			$score(r) \leftarrow \min \{ {score_1},{score_2},...,{score_{w}}\} $\\
    		}
		// \ding{188}\textsc{Score Averaging}:\\
		$ Score \leftarrow \frac{1}{R}\sum\limits_{r = 1}^R {score(r)} $\\
		\Return{$Score$}
\end{algorithm}

\begin{algorithm}
\SetAlgoLined
\DontPrintSemicolon
	\caption{Jenkins Hash Function}
	\label{al:jenkins}
		\KwIn{The string: $key$, string length: $len$ and random seed: $seed$}
		\KwOut{The hash code of input string.}
		${\rm{hash}} \leftarrow {\rm{seed;}}$\\
	    \For {$i=1,2,\ldots,len$}{
	        \#pragma HLS PIPELINE\\
            $hash \leftarrow hash + key[i]$\\
            $hash \leftarrow hash + (hash <  < 10)$\\
            $hash \leftarrow hash \oplus (hash >  > 6)$\\
        }
        
        $hash \leftarrow hash + (hash <  < 3)$\\
        $hash \leftarrow hash \oplus (hash >  > 11)$\\
        $hash \leftarrow hash + (hash <  < 15)$\\
        $hash\_code \leftarrow hash{\rm{ }}\% {\rm{ }}MOD$\\
		\Return{$hash\_code$}\\
\end{algorithm}

The purpose of \textit{projection} is to reduce the dimensionality of a set of points, whilst retaining the essence of the original data. Randomness between sub-detectors to improve diversity in an ensemble is also introduced at this step. It is also the most computationally expensive step.
The \textit{core} block is the cornerstone of each method: Loda is based on the histogram; RS-Hash and xStream approaches make use of the count-min sketch (CMS)~\cite{cms2005}, in which \textit{w} pair-wise independent hash tables are used. These three methods all operate over a \textit{sliding-window} of the input data which is received in a streaming fashion. 
The difference is that the histogram-based Loda only uses a 1-row table as a sliding-window, where the CMS based methods allow the sliding-window with a \(w\)-row table ($w\ge1$). The \textit{score} block calculates the negative log-likelihood so the less likely a sample, the higher the anomaly value.

Algorithm~\ref{al:loda} to Algorithm~\ref{al:xstream} present the pseudo-code for Loda, RS-Hash and xStream respectively. Algorithm~\ref{al:jenkins} introduces the hash function which is applied in RS-Hash and xStream. Clearly, each function has the streaming input \(x\) and the streaming output \(Score\). 

The ensemble can be divided into seven parts. The \ding{182}\textsc{Windower} block uses a shift register to assemble samples ${x} \in \mathbb{R}^1$ into an entire vector $\vec{x}_i = \{{x}_1,{x}_2,...,{x}_d\} \in \mathbb{R}^d$, \ding{183}\textsc{Ensemble} is a big \textit{for} loop with \(R\) independent iterations. Each iteration can be regarded as a sub-detector with a unit of 1, and each sub-detector executes the functional modules of \ding{184}\textsc{Projection}, \ding{185}\textsc{Core} (Histogram for Loda, and Hash Function for RS-Hash and xStream) \ding{186}\textsc{Sliding-window} and \ding{187}\textsc{Score} sequentially. The execution of these modules is time-dependent, that is, the start of the latter one must wait for the execution of the former to complete. Finally, \ding{188}\textsc{Score Averaging} is used for producing the final ensemble anomaly score by averaging the outputs of all sub-detectors.

The serial combination of these four blocks (\ding{184}\ding{185}\ding{186}\ding{187}) constitutes a base sub-detector. We use \(R\) parallel executions of the base sub-detector with a different starting seed or hash and average their scores, to form an \textit{ensemble}. For a CPU implementation, \(R\) sub-detectors are processed sequentially. Details of HLS directives used in the pseudo-code are described in Section~\ref{se:3_module_generator}.

\subsection{Ensembles of multiple anomaly detectors}
\label{se:2_combo}

Ensembles are a class of methods that combine weak detectors to collectively form a more accurate decision by utilising diversity in the detectors~\cite{Freund1997,Freund1996}. Each sub-detector in an ensemble is data-independent and structure-identical. The first feature makes it naturally amendable for parallel processing, while the latter gives it the flexibility to assemble arbitrary numbers of sub-detectors into an ensemble according to the computational resources and desired accuracy. 

\begin{figure}
\centerline{\includegraphics[width=0.6\linewidth]{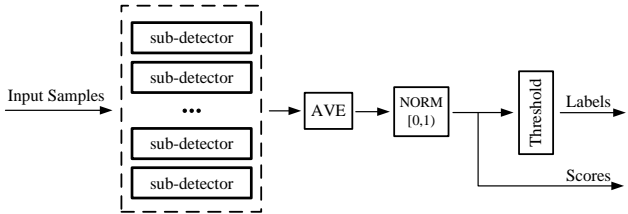}}
\caption{Sample Architecture for SEAD Methods.}
\label{fg:sead_architecture}
\end{figure}

An example architecture for streaming ensemble anomaly detectors (SEADs) is illustrated in Figure~\ref{fg:sead_architecture}, where each sub-detector inside SEAD takes a stream of input data and produces a stream of transformed outputs which indicate the anomaly scores. Averaging is used to compute the final score from all sub-detectors, or a \textit{threshold} can be applied to translate this averaged score to a \textit{Labels} (anomaly or no anomaly). While a simple ensemble could be multiple instances of the same detector, as described in the previous section, a more powerful ensemble will utilise different detectors.

An easy-to-use scalable library provides the opportunity to explore the performance of ensembles, many of which have been developed. Examples of anomaly detection packages include: ELKI Data Mining~\cite{ELKI2010} and RapidMiner~\cite{RapidMiner2016} in Java; Outliers~\cite{R_outlier2011} in R; SUOD~\cite{suod2021}, PyOD~\cite{pyod2019} and PySAD~\cite{pysad2020} in Python. Aside for their differences in programming languages, different libraries are also tailored to different kinds of anomaly detection, e.g. PySAD focus in particular on a framework of streaming ADs in Python, whereas only static approaches can be accessed by SUOD and PyOD.

In addition to supporting comprehensive detectors, the libraries provide the ability to combine the output of these models in different ways beyond simply taking the average or maximum across all the base models. An software toolkit \(combo\)~\cite{combo2020} contains more than 15 model combination methods in Python, including basic generic and global methods (like Averaging, Maximization, Weighted Averaging, Feature Bagging etc.\cite{ensemble_Zimek2014,Feature_bagging2005,theroy_ensemble_2015}) and dynamic selection/combination models (such as DCS~\cite{DCS1994} and LSCP~\cite{LSCP2019} etc.). While our library currently only performs averaging inside an ensemble, other combination techniques are easily implemented.

\subsection{Dynamic Partial Reconfiguration and Partial Overlays}
\label{se:2_partial_reconfiguration}
FPGA technology provides the flexibility to modify a hardware implementation without re-fabrication.
Partial reconfiguration (PR) takes this flexibility a step further, allowing dynamic changes to an active design.
This requires implementing static logic, multiple Reconfigurable Partitions (RPs) with various Reconfigurable Modules (RMs). The RP is the level of hierarchy within which different RMs can be implemented and an RM is the netlist or HDL description that is implemented within an RP. Generally, Multiple RMs with the same interface exists for a specific RP. 
Dynamic Function eXchange refers to a Xilinx tool flow that achieves the partial reconfiguration~\cite{DFX,DFX_tutorial}.

\begin{figure}
\centerline{\includegraphics[width=0.7\linewidth]{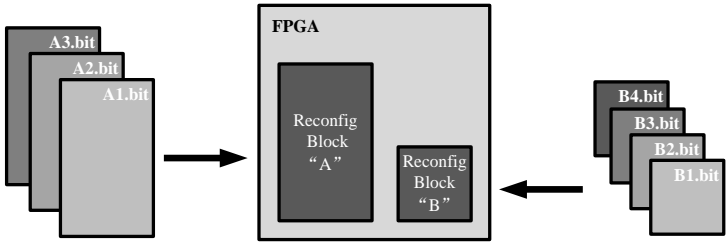}}
\caption{Basic Premise of Partial Reconfiguration.}
\label{fg:partial_reconfiguration}
\end{figure}

Figure~\ref{fg:partial_reconfiguration} illustrates the basic premise of partial reconfiguration. The grey area of the FPGA block represents static logic, and the block marked as Reconfig Block `A' and Reconfig Block `B' represent the RPs. The functionality implemented in Reconfig Block `A' can be modified by downloading one of several partial BIT files: A1.bit, A2.bit or A3.bit; similarly the functions implemented in Reconfig Block `B' can be modified by one of B1.bit to B4.bit.

Using DFX allows the fSEAD library to be able to support improvements or new ADs developed in the future. Moreover, supporting multiple customized partial regions enables  fSEAD to support ADs that have different resource requirements. This flexibility is also important because different applications will have different hardware capabilities and accuracy requirements, while the optimal performance may change depending on external conditions, such as environmental changes.

\subsection{Literature Review}
\label{se:2_literature_review}
This section reviews other anomaly detection techniques, literature on FPGA-based anomaly detection, and briefly highlights related research in dynamic reconfigurable FPGA implementations for other applications. 
A powerful class of anomaly detectors utilize tree-based structures, with examples including iForest~\cite{iForest2008}, HS-Tree~\cite{HSTree2011}, and RS-Forest~\cite{RSForest2014}. The Isolation Forest (iForest) approach builds an ensemble of ``Isolation Trees'' (iTrees) for the dataset, and anomalies are the points that have shorter average path lengths on the iTrees. HS-trees, are similar to Isolation Forest, but decision rules within tree-nodes are generated randomly. RS-Forest takes a further step by using multiple fully randomized space trees to tackle the streaming detection problem from the density estimation aspect, which is more efficient as it leverages the common operations shared by the prediction and model update processes.

Kernel methods and data-centred models have also been proposed as online anomaly detectors. Support Vector Regression (SVR) with Gaussian kernel-based online novelty detection on temporal sequences is presented by Ma et al.~\cite{kernal2007}. High-performance FPGA implementations of online kernel methods was demonstrated in references~\cite{krls2013,fastfood2013}. Based on static data-centred LOF methods~\cite{LOF2000}, an incremental LOF algorithm, appropriate for detecting outliers in data streams is proposed in~\cite{LOF_online2007} which provides equivalent detection performance as the iterated static LOF algorithm, while requiring significantly less computational time. Das et al. proposed a system based on feature extraction and Principal Component Analysis (PCA)~\cite{pca_fpga2008}. Their FPGA implementation could support data at over 20 Gbps. Pang et al. \cite{krls2013} present a high-performance FPGA implementation that achieves improvements in execution time, latency and energy consumption by factors of 5, 5 and 12 respectively over CPU and digital signal processor (DSP) implementations for the online kernel method~\cite{kernal2007}. Hayashi and Matsutani \cite{niclof2017} offload the Local Outlier Factor (LOF) calculation to a FPGA based Network Interface Card for online anomaly filtering. This leads to throughput improvements of up to 10x on the anomaly filtering over a software-based execution.

Neural Network models have also been proposed for streaming anomaly detection. Moss et al. \cite{autoencoder_fpga2018} introduce an FPGA accelerated Neural Network-based anomaly detector based on an auto-encoder for processing of physical-layer radio-frequency (RF) signals. This design processes continuous 200 MS/s complex inputs, producing anomaly classifications at the same rate, with a latency of 105 ns, an improvement of at least 4 orders of magnitude over a conventional approach using a software defined radio.

While there are published FPGA implementations of random  forests~\cite{RForest_cpugpufpga2012,RForest_HLS2018,RForest_Reconfig2017},  kernel and neural network based FPGA-based accelerators, these are all fully-customized hardware designs for a specific algorithm. Moreover, they typically utilise most of the available FPGA resources meaning it is difficult to implement such detectors in a partial region of an FPGA. The focus of this article is on the development of a flexible library. Unlike the many comprehensive and easy-to-use anomaly detection libraries released on different software platforms~\cite{ELKI2010,RapidMiner2016,R_outlier2011,suod2021,pyod2019,pysad2020}, we believe this is the first flexible FPGA library for anomaly detection in the literature.

Although we are not aware of any publications describing runtime reconfigurable anomaly detection libraries, dynamic reconfiguration on FPGAs has been used for other FPGA applications. The most similar to this work is Wilson who proposed a real-time video processing pipeline that utilizes the dynamic reconfigurable aspects of FPGA ~\cite{videopipeline2020}. Their work used 11 reconfigurable regions, allowing for multiple custom run-time configurations and it adopts the partial bitstreams with PYNQ overlays to ease the software development. Our work is different in that it uses AXI switches to dynamically switch anomaly detectors and model combinations between reconfigurable regions, and employ a flexible module generator to create the regions themselves.

\section{Design}
\label{se:3_fsead_methodology}
In this section we first provide a high-level description of the fSEAD system, followed by the module generator for creating integrated anomaly detection IPs. Next we introduce the DFX workflow we abstracted for creating partial reconfiguration project in Xilinx Vivado environment. We then describe the interconnection scheme which provides a high degree of flexibility and enables IP modules to be composed at run-time, allowing applications to be efficiently accelerated without regeneration of bitstreams. Finally the FPGA implementation is discussed.

\begin{figure}
\centerline{\includegraphics[width=1.0\linewidth]{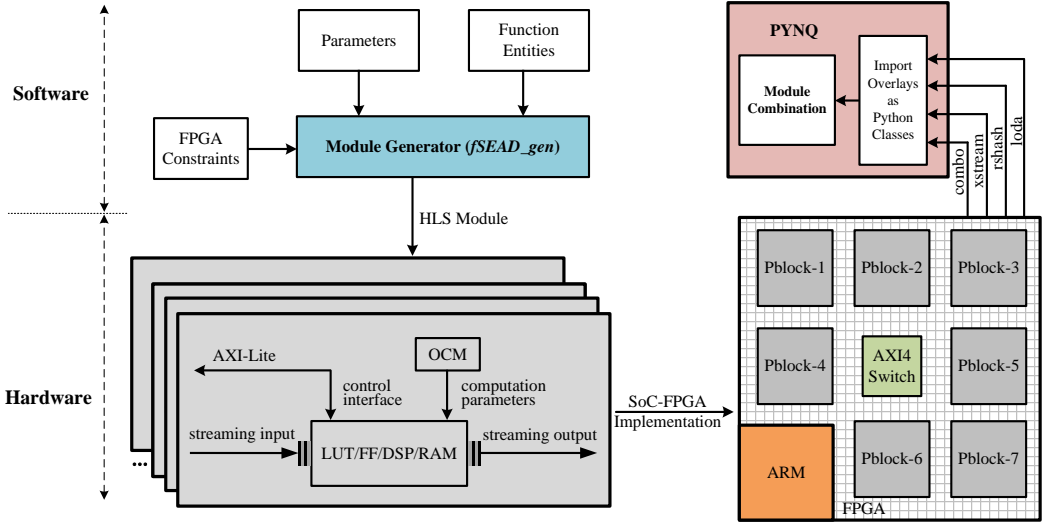}}
\caption{Overview of the fSEAD Framework. }
\label{fg:fsead_framework}
\end{figure}

The system framework is illustrated in Figure~\ref{fg:fsead_framework}. It consists of four software and hardware components.
\textit{fSEAD\_gen} in the upper-left corner of Figure~\ref{fg:fsead_framework} is a Python-based module generator. It takes parameterised function entities and produces Vivado HLS modules under. 
The lower-left corner of Figure~\ref{fg:fsead_framework} shows the interface for pblocks, which take streaming inputs and produce streaming outputs. These then are encapsulated as multiple unique IPs and wrapped within AXI streaming interfaces and an AXI-Lite controller. All parameters required by the IP modules are stored in on-chip memory (OCM) for performance. 
The lower-right sub-figure shows multiple pblock regions. Multiple sub-detector IPs are arranged in a spatially parallel fashion within each pblock.  Each pblock is implemented by many reconfigurable modules (RMs) and can be customised at run-time.

\subsection{Module Generator}
\label{se:3_module_generator}
The module generator allows customisation of the underlying sub-detectors for latency optimization and resource utilisation exploration.

The \textit{fSEAD\_gen} module generator, written in Python, takes as inputs: the anomaly detector parameters, data-type, precision, function description, target dataset and a testing set. It produces a standalone C program suitable for synthesis via HLS as output. The parameters, interface, and optimization directives are all embedded in the C program.
A compact sub-detector C instance is formed by combining the \ding{184}\textsc{Projection}, \ding{185}\textsc{Core}, \ding{186}\textsc{Sliding-window} and \ding{187}\textsc{Score} parts of Algorithm~\ref{al:loda}-\ref{al:xstream}. 
Arbitrary numbers of sub-detectors, specified by the user to the module generator, are integrated in parallel manner to form an ensemble.
A self-verifying test-bench compares the program with golden results from the original Python description. Thus programming errors introduced by the generated ensemble program can be quickly identified in the simulation step. For synthesis, compiler directives such as DATAFLOW, ARRAY PARTITION, UNROLL and PIPELINE are used to aid translation of the C description to a spatially parallel RTL circuit.

In Algorithm~\ref{al:loda} to Algorithm~\ref{al:xstream}, we use the DATAFLOW pragma only in the top layer (line-1 of each algorithm) to enable task level pipelining. This allows functions and loops to execute concurrently and achieve sub-detector parallelism.
UNROLL is used to create multiple instances of the loop body in the RS-Hash and xStream algorithms to exploit spatial parallelism.  This enables all hash functions inside the CMS table (line-16 in Algorithm~\ref{al:rshash} and line-18 in Algorithm~\ref{al:xstream}) to execute concurrently, and similarly accelerates the  projection of size $K$ for xStream (line-12 in Algorithm~\ref{al:xstream}).
PIPELINE reduces the initiation interval ($II$, the number of clock cycles before the function can accept new input data) by allowing overlapped execution of operations within a loop or function. It is used in the $projection$ loop body of all detectors (line-10 of Algorithms~\ref{al:loda} to ~\ref{al:xstream}) and the $for$ loop in the Jenkins hash function (line-3 of Algorithm~\ref{al:jenkins}). $II=1$ is achieved in all the loop bodies with PIPELINE for our implementation. 
The result of all these optimisations is a highly parallel design which balances resource utilization and latency.

\begin{figure}
\centerline{\includegraphics[width=1.0\linewidth]{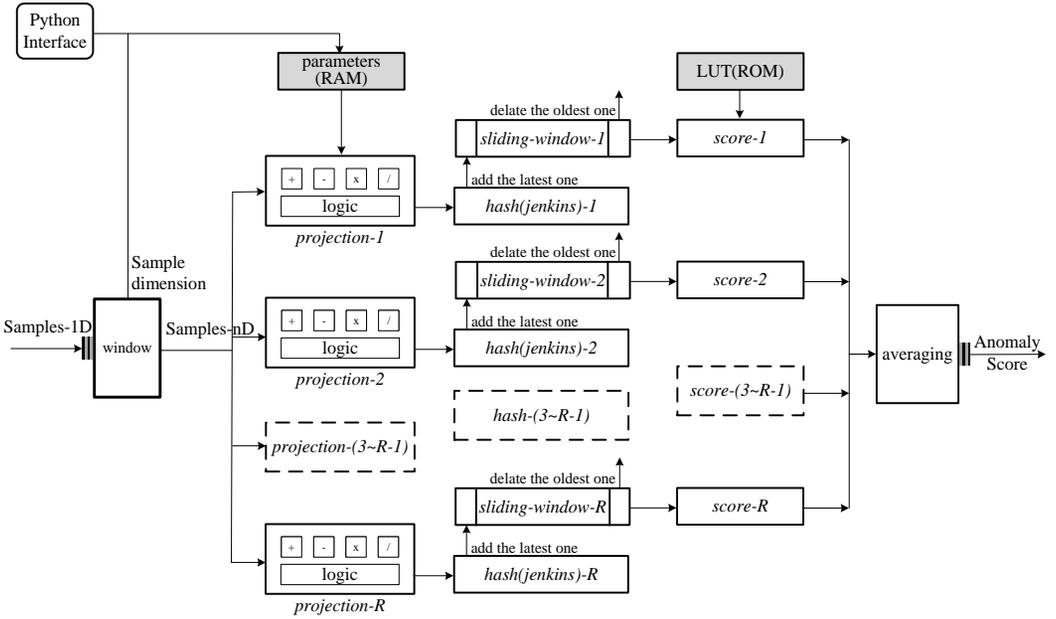}}
\caption{An Example of A Hash-based AD Hardware Structure in fSEAD.}
\label{fg:sead_circuit}
\end{figure}

\textit{fSEAD\_gen} currently supports three types of detectors with arbitrary ensemble size and coefficients (Loda, RS-Hash and xStream). All HLS directives are first manually designed for the target FPGA and then embedded in \textit{fSEAD\_gen}. \textit{fSEAD\_gen} can automatically generate optimizations for different ensemble sizes, such as hyper-parameters and other configurations, without modifying the existing HLS directives. However, developers maintain the flexibility to adapt the current HLS directive to find more  optimized solutions for a specific area-latency trade-off requirement. New detectors can be written in C/Python and easily integrated using existing detectors as examples. In this case, specific optimized HLS directives have to be manually re-designed for new members to the fSEAD library.

Figure~\ref{fg:sead_circuit} shows an ensemble of hash-based detectors which have a more complex hardware implementation than Loda. 
For simplicity, this figure highlights the parallel sub-detectors generated by the DATAFLOW pragma, instead of the detailed circuits from the lower level by PIPELINE and UNROLL pragmas. This represents the main acceleration of fSEAD: task-level parallelism for sub-detectors operating concurrently.

Input data is streamed into the IP on the left interface and outputs the real-time anomaly score on the right side. Input samples are windowed to assemble a batch with target dimension: \(d\) (refer to \ding{182}\textsc{Windower} in Algorithm~\ref{al:rshash} and Algorithm~\ref{al:xstream}). Computation of all the streams occur concurrently, before a final reduction step (refer to \ding{183}\textsc{Ensemble}, in Figure~\ref{fg:sead_circuit} this is averaging). Sub-detector level parallelism is achieved by applying the HLS DATAFLOW directive on the top function. In this way, the maximum ensemble size is only limited by resources available in the target FPGA. Inside each sub-detector, small area and low latency are the goals. We compute each row of a CMS table using independent hash functions in parallel (refer to \ding{185}\textsc{Hash-Function}). PARTITION directives are used to divide a large RAM into smaller storage units, increasing available parallelism of accesses. This is advantageous because even with dual-port RAM, only one read and one write operation can be completed in one clock period. Notably, since the usage of PARTITION significantly affects the resource consumption of LUTs or FFs on FPGA, it is only used for a specific dimension of the target array. This guarantees the functions of DATAFLOW, PIPELINE and UNROLL directives. All HLS directives above enable a balanced optimization inside an ensemble AD instance, which is independent to the partial-block-level model combination scheme introduced in the following Section~\ref{se:3_composable}.

To avoid high resource cost with little throughput benefit, we constrain the \ding{184}\textsc{Projection} module using PIPELINE instead of UNROLL inside each sub-detector.
To compute the logarithm required for the negative log-likelihood score of Table~\ref{tb:sead_methods}, a $W$-deep lookup table with 32-bit representation is used for window-size of $W$.

\subsection{DFX Tool Flow}
\label{se:3_DFX_toolflow}

\begin{figure}
\centerline{\includegraphics[width=1.0\linewidth]{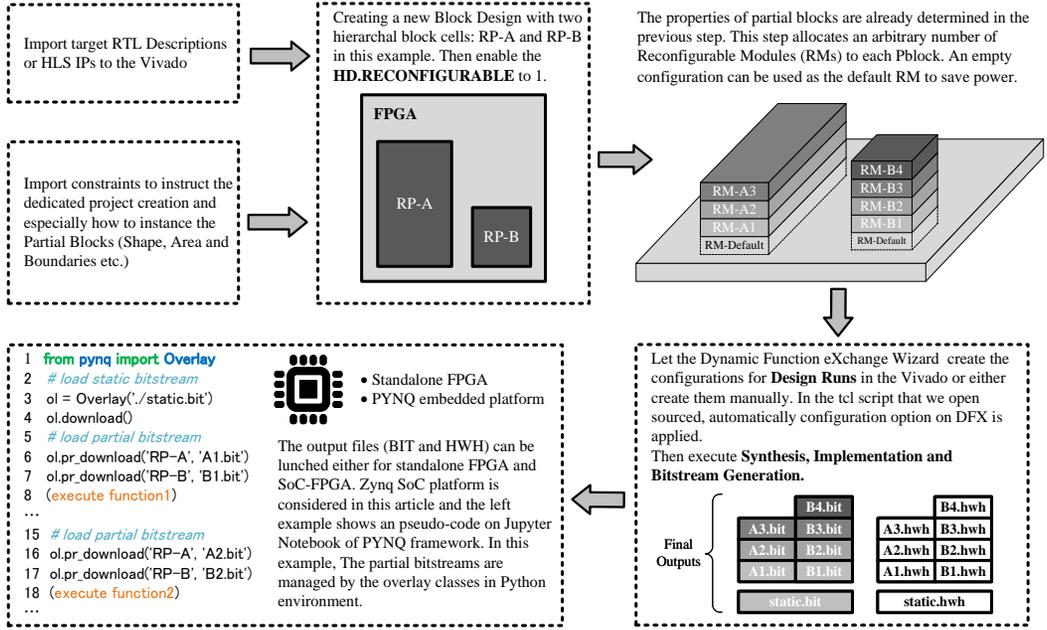}}
\caption{An Example of Xilinx Partial Reconfiguration Tool flow.}
\label{fg:dfx_toolflow}
\end{figure}

DFX is an important tool to endow fSEAD with partial reconfigurability. Its customized workflow for fSEAD is introduced in this section. Continuing on the basis of the example in Figure~\ref{fg:partial_reconfiguration}. Figure~\ref{fg:dfx_toolflow} shows an example of how two reconfigurable areas are mapped to Reconfigurable Partitions (RPs) using our tool flow.
It should be noted that in addition to the dynamic RMs for each Partial block (e.g. the RM-A1, to RM-A3 for RP-A and RM-B1 to RM-B4 for RP-B), a default RM can also be assigned for each pblock. The logic of the RM-Default will be first active when the static.bit is downloaded, which brings with a recommended way of setting the default RM to an empty logic to save power before this pblock is configured by the dedicated RM of users.

In the last step (the lower left sub-figure) in Figure~\ref{fg:dfx_toolflow}, we show that standalone FPGA or Xilinx PYNQ platforms are able to support the fSEAD execution. We select the PYNQ as the final execution platform in this example. 
The bitstreams and a hardware hand-off (HWH) file are used as overlays by PYNQ to automatically identify the Zynq system configuration, IP including static regions and partial blocks.

\subsection{Composable Infrastructure}
\label{se:3_composable}
The Xilinx DFX tool~\cite{DFX} supports partial reconfiguration of RMs in an FPGA  while the rest of the device remains operational. There is no limitation in the number of RMs supported. In fSEAD, each pblock is a unique RM configuration with its own BIT file. By downloading the appropriate one, the hardware functionality can be customised at run-time.  In summary, pblocks are either: (1) Ensembles of homogeneous sub-detectors implemented in a pblock; and (2) Combination modules which aggregate heterogeneous pblock output streams. 

An AXI4 switch acts as a router and orchestrates data movements between pblocks. Inputs can be routed from multiple input streams to any pblock, with outputs routed to remaining pblocks and back to a host processor.

\begin{figure}
\centerline{\includegraphics[width=1.0\linewidth]{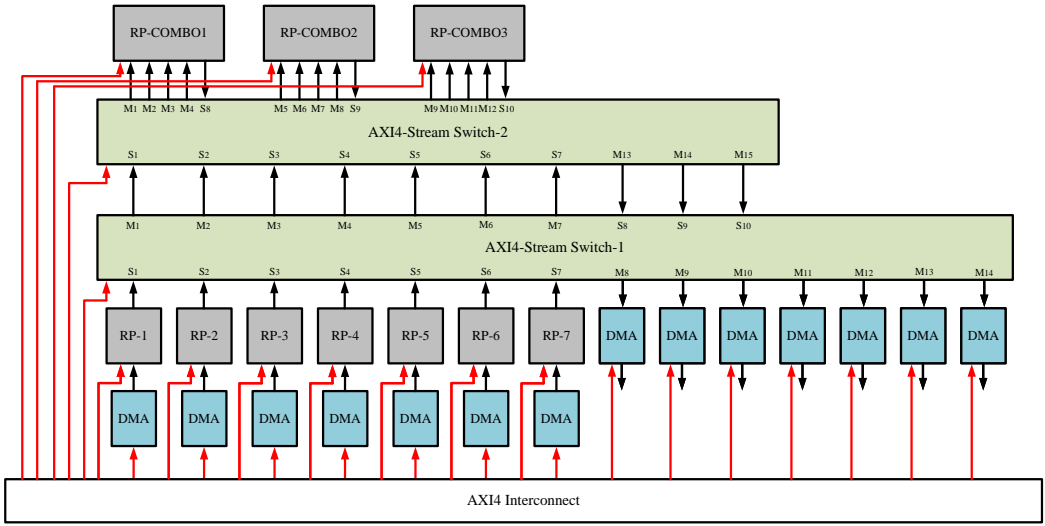}}
\caption{Composable Topology.}
\label{fg:composable_topology}
\end{figure}

Figure~\ref{fg:composable_topology} shows the composable topology proposed in fSEAD. The grey regions are pblocks. Blue blocks symbolize Direct Memory Access (DMA) controllers for data transfer. Two AXI4-Stream switches~\cite{xilinx_Stream} enable dynamic routing from a Slave port to a Master port, e.g. for switch-1, S1 in the lower left is the first Slave interface and M14 in the lower right symbolizes the 14th Master interface. Master and Slave interfaces are symmetric and point-to-point, so that Master output signals can connect directly to Slave input signals. Any number of external modules can be daisy-chained together. The modules can be used for a multitude of different tasks such as buffering, data transform or routing. Multiple switches are used since each Xilinx AXI4-Stream Switch IP switch only supports a maximum of 16 Slave ports and 16 Master ports. Cascades of two or more switches allow an arbitrary number of pblocks to be interconnected. The black lines depict AXI4-Stream connections, and the red lines are AXI-Lite bus connections. These are connected and controlled by the AXI4 Interconnect at the bottom.

In our prototype implementation, seven independent pblocks are available for implementing anomaly detectors (shown as RP-1 to RP-7 in Figure~\ref{fg:composable_topology}. Each pblock has one AXI4-Stream input connected to a fixed DMA and one output interface connected to a Slave port of AXI4-Stream Switch-1. Multiple sub-detectors can be placed in a single pblock, each with different parameter settings. For example, 35 Loda sub-detectors fit in a single pblock. The combo pblocks at the top of Figure~\ref{fg:composable_topology} are responsible for aggregation. Each is equipped with four input ports and one output port, all connected to Switch-2. 

Table~\ref{tb:combo_score_label} shows the currently supported combination methods implemented in fSEAD for continuous (score) and discrete (label) targets.  The label targets are `0' meaning not an anomaly and `1' for anomaly.
General and global (GG) methods~\cite{LSCP2019,ensemble_Zimek2014,Feature_bagging2005,theroy_ensemble_2015} are used by popular machine learning libraries such as scikit-learn~\cite{scikit}, XGBoost~\cite{XGBoost} and LightGBM~\cite{lightgbm} for aggregation of sub-detector class results. In fSEAD, for sub-detectors with continuous outputs we implemented three representative GG methods, namely Averaging (\textit{GG\_A}), Maximum (\textit{GG\_M}), and Weighted Average (\textit{GG\_WA}). 
To combine label outputs, two commonly used approaches: or (a class is true if any sub-detector outputs true for that label) and voting (the class with most votes is chosen as the output) are applied.
In this work, we always use averaging for combining anomaly scores; and the or-gate technique to combine labels. Of course, this can be customised by the user.

\begin{figure}
\centerline{\includegraphics[width=0.95\linewidth]{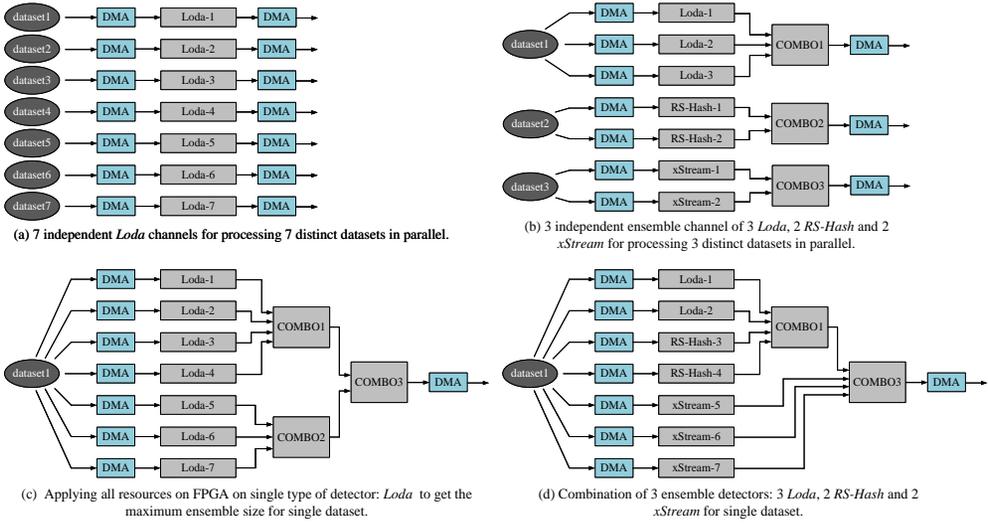}}
\caption{Examples of Combination Scheme.}
\label{fg:combination_scheme}
\end{figure}

Routing through the AXI switches is configured via the AXI-Lite interface. A register is used for each master to slave connection. After the registers have been configured, the interconnection is determined. Unused master or slave interfaces are disabled. When a slave interface is connected to multiple masters, only the lowest numbered one is used, e.g. if both Master-1 and Master-3 are configured to connect to Slave-2, then Master-1 wins the arbitration and Master-3 is disabled. Thus effectively only one connection between each master and slave is made. 

Figure~\ref{fg:combination_scheme} shows four example configurations of our composable topology. Figure~\ref{fg:combination_scheme}(a) shows the simplest case where  seven parallel Loda pblocks (each containing multiple sub-detectors) are used to analyse seven different datasets. Each streaming channel implements a unique and independent anomaly detection application. Switch-1 is configured to directly route RP-1 to RP-7 to seven output DMA channels. This configuration only requires Switch-1 so connections to Switch-2 and the combo pblocks are disabled.
The case in Figure~\ref{fg:combination_scheme}(b) implements three independent anomaly detection applications. It exploits all pblocks and two of the Switches. RP-1, RP-2 and RP-3 implement a Loda ensemble and their outputs are routed to the inputs of COMBO1. The output is the final score which is sent to the host via DMA. The pblocks (RP-4, RP-5) and (RP-6, RP-7) generate RS-Hash and xStream ensembles for two different datasets. 
Figure~\ref{fg:combination_scheme}(c) is an example that only operates on a single dataset and a single type of anomaly detector, namely Loda. It uses all the pblocks to implement a maximally parallel, homogeneous ensemble.
Finally, Figure~\ref{fg:combination_scheme}(d) is similar to Figure~\ref{fg:combination_scheme}(c) but incorporates three different types of detectors: Loda, RS-Hash and xStream.

The configurations are not limited to the four cases just described. By providing different pblocks and switch configurations, a multitude of customized anomaly detection functionalities can be implemented in a run-time configurable manner.

\subsection{FPGA Implementation}
\label{se:3_fpga_implementation}
This section continues the example of Figure~\ref{fg:combination_scheme}(c) as concrete example to describe physical implementation on a Zynq UltraScale+ ZCU111 Evaluation Board with XCZU28DR-2FFVG1517E RFSoC FPGA.

\begin{table}
  \caption{Combination Methods for Scores and Labels}
  \scalebox{0.8}{
  \begin{tabular}{lcl}
    \toprule
    Output & Combination Methods & Equation\\
    \midrule
    score & Averaging & $combo = (scor{e_1} + scor{e_2} +  \cdots  + scor{e_N})/N$\\
    
    score & Maximization & $combo = max(scor{e_1},scor{e_2}, \cdots ,scor{e_N})$\\
    
    score & Weighted Average & $combo = ({w_1} * scor{e_1} + {w_2} * scor{e_2} +  \cdots  + {w_N} * scor{e_N})/N,(\sum\limits_{i = 1}^N {{w_i}}  = 1)$\\
    
    label & Or & $combo = (labe{l_1}|labe{l_2}| \cdots |labe{l_N})$\\
    
    label & Voting & $combo = voting(labe{l_1},labe{l_2}, \cdots ,labe{l_N})$\\
  \bottomrule
\end{tabular}}
\label{tb:combo_score_label}
\end{table}

\begin{figure}
	\centering
	\begin{minipage}{0.49\linewidth}
		\centering
		\includegraphics[width=0.9\linewidth]{./figures_tables/FPGA_layout.pdf}
		\caption{FPGA Layout.}
		\label{fg:fpga_layout}
	\end{minipage}
	\begin{minipage}{0.49\linewidth}
		\centering
		\includegraphics[width=0.9\linewidth]{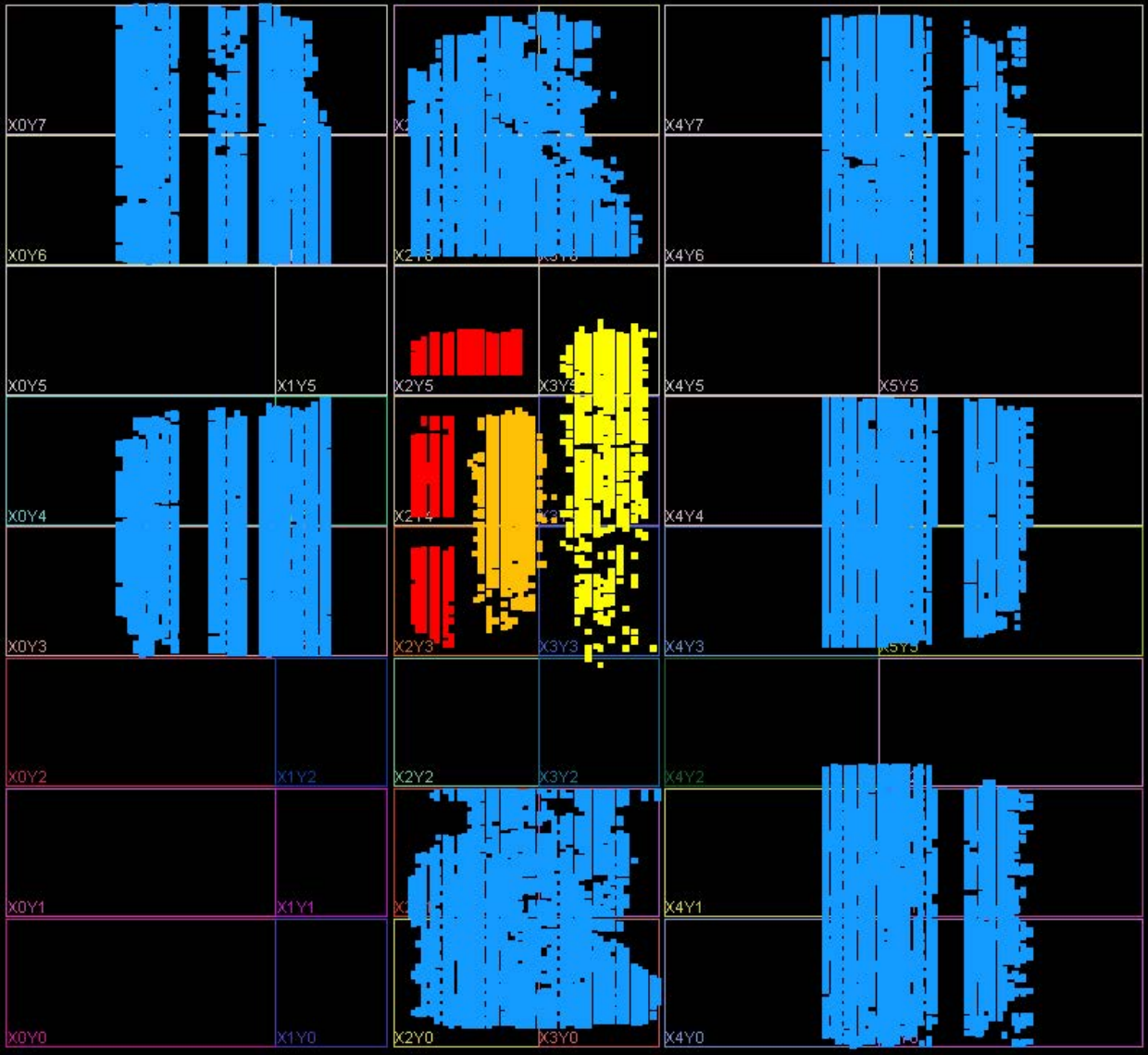}
		\caption{Placement on FPGA.}
		\label{fg:fpga_placement}
	\end{minipage}
\end{figure}

Figure~\ref{fg:fpga_layout} and Figure~\ref{fg:fpga_placement} show the FPGA layout and placement respectively. The floorplan is based on two considerations: (1) Seven AD-pblocks occupy the main FPGA resources to ensure that the fSEAD has sufficient resources to implement parallel ensembles and  complex detectors. In comparison, the combo modules, switches and DMA units are not resource-intensive and are reflected in the layout; (2) Although long interconnections are needed because of the distributed pblocks, this is alleviated through AXI bus-level pipelining~\cite{xilinx_Interconnect}. This technique isolates the path between the master and slave with registers, while maintaining an AXI4 protocol compliant pipeline stage. The register slice is implemented as a two-deep FIFO buffer that supports bus communication without generating unnecessary idle cycles. It serves to achieve timing closure by reducing the critical path.

We place Switch-1 (yellow) in the centre of the FPGA with Switch-2 (orange) adjacent since they two communicate directly with each other. Switch-1 is assigned a larger area than the Switch-2 as it connects to seven pblocks (in blue), while Switch-2 is smaller to prevent blocking routes from RP-1 and RP-4 to Switch-1. As shown in the Figure~\ref{fg:fpga_layout}, the nearest routing channel from RP-1 to Switch-1 is only a very narrow programmable slot (white line). The three combo-pblocks (red) only connect to Switch-2. The floorplan for the above-mentioned blocks is compact and sits in the middle layout of FPGA.

The seven AD-pblocks (in blue) occupy the remaining resources. It is important to note that we did not issue placement area constraints for DMAs, AXI4-Interconnect and the other static components. Instead, we allow these components to be placed by the Vivado tool to minimizing routing delay. The spaces between the coloured regions are available for any remaining static logic. Furthermore, during the Partial Reconfiguration process, the DFX Decoupler~\cite{DFX} is used to allow users to isolate the logic being configured until it is done and the new  logic reset. 

\section{Results}
\label{se:4_experiment}
The aforementioned techniques were implemented and results presented in this section. In Section~\ref{se:4_test_platform} we introduce the development environment and test platform.
In Section~\ref{se:4_ensemble} we demonstrate why larger ensembles with more sub-detectors is desirable, and how performance is boosted in terms of accuracy from model combination.
We then discuss the resource use of the static regions, and available resources in each of the FPGA partial regions in Section~\ref{se:4_partial_blocks}. 
Furthermore, we show the performance gains of our architecture over a CPU in Section~\ref{se:4_latency_accuracy}.
Finally in Section~\ref{se:4_dfx_speed_generic_useage}, we analyse the performance characteristics of partial reconfiguration  and more general features of the fSEAD architecture.

\subsection{Test Platform}
\label{se:4_test_platform}
We utilize the Xilinx Zynq UltraScale+ ZCU111 (xczu28dr-2ffvg1517e) board for evaluating the fSEAD library. \textit{fSEAD\_gen} is used to generate three anomaly detectors (Loda, RS-Hash and xStream) with  HLS and GCC targets used for FPGA and multi-threaded CPU implementation respectively. The HLS module is synthesised to RTL using the Xilinx Vivado HLS tool (v2020.1) and then passed through Xilinx Vivado Design Suite (v2020.1). We then deploy the architecture on the FPGA using the PYNQ framework. The GCC compiled versions of Loda, RS-Hash and xStream as CPU benchmarks are tested on desktop with Intel(R) Core(TM) i7-10700F @2.9GHz and 64GB memory for performance comparisons. The g++ compiler is used with flags `-Wall --std=c++14 -g -O3', with `-lpthread' applied for multi-threaded optimization. The flag `-ftree-vectorize' is turned on by default under `-O3' to enable automatic vectorization for further optimization.

The anomaly detection performance evaluation was conducted on four publically available datasets: Cardio, Shuttle, HTTP-3 and SMTP-3, with their main attributes summarised in Table~\ref{tb:datasets}. Cardio and Shuttle are also used in SUOD~\cite{suod2021}. HTTP-3 and SMTP-3 were derived from the KDD-cup 99~\cite{kddcup99} dataset and are 3 feature variants of the 41 feature full datasets, first used in~\cite{HSTree2011}. The sample number \(n\) varies from 1831 (Cardio) to 567498 (HTTP-3) and the dimensionally ranges from 3 to 21. Cardio proportionally has the largest number of anomalies (9.61\%), and only 0.03\% of samples in SMTP-3 were in the anomaly category. In addition to the four datasets above, others are available including those in the ODDS Library~\cite{ODDS}, UCI repository~\cite{UCI} and DAMI Datasets~\cite{DAMI} for future research. 

Accordingly to the SEAD structure in Figure~\ref{fg:sead_architecture}, the output scores of each detector are first normalized to [0,1). A sample with a higher score indicates a higher probability that this sample belongs to the anomaly category. Furthermore, with the anomaly percentage, or named contamination rate that the users know in advance, a threshold can be determined to translate anomaly scores to binarized anomaly labels. For example, if the threshold is 0.8, then inputs with anomaly scores of 0.9 and 0.5 would be assigned to labels `1' (anomaly) and `0' (normal) respectively. 
The standard metric used for evaluating anomaly detectors is Area Under the Curve (AUC) of the Receiver Operating Characteristics (ROC) curve. It can be used for scores or labels and is described in detail in ~\cite{AUC}. We adopt this metric to analyse the accuracy of our implementations.

\begin{table}
  \caption{Datasets}
  \scalebox{0.8}{
  \begin{tabular}{lrrrr}
    \toprule
    Datasets & Sample Length & Dimension & Outliers & \%Outliers\\
    \midrule
    Cardio & 1831 & 21 & 176 & 9.61\\
    Shuttle & 49097 & 9 & 3511 & 7.15\\
    SMTP-3 & 95156 & 3 & 30 & 0.03\\
    HTTP-3 & 567498 & 3 & 2211 & 0.40\\
  \bottomrule
\end{tabular}}
\label{tb:datasets}
\end{table}

\subsection{Sub-detectors and Ensemble Accuracy}
\label{se:4_ensemble}
We conducted an experiment to show the effectiveness of increasing the number of sub-detectors for each of the three supported methods in fSEAD.

For different ensemble sizes, the AUC of detectors in fSEAD and the variance of AUC are evaluated over 10 executions with different random seeds. The hyper-parameters were set to the values in Table~\ref{tb:hyper_param}, these being chosen to give an accurate and efficient hardware implementation. We refer the reader to the relevant papers regarding selecting these hyper-parameters~\cite{loda2016,rshash2016,xstream2018}.

\begin{table}
  \caption{Hyper-Parameter Set for the Three Detectors}
  \scalebox{0.8}{
  \begin{tabular}{lccccc}
    \toprule
    Detector & \textit{window size} & \textit{Bins} & \textit{CMS-w} & \textit{CMS-MOD} & \textit{K}\\
    \midrule
    Loda & 128 & 20 & 1 & - & -\\
    RS-Hash & 128 & - & 2 & 128 & -\\
    xStream & 128 & - & 2 & 128 & 20\\
  \bottomrule
\end{tabular}}

\label{tb:hyper_param}
\end{table}

\begin{figure}
\centerline{\includegraphics[width=0.7\linewidth]{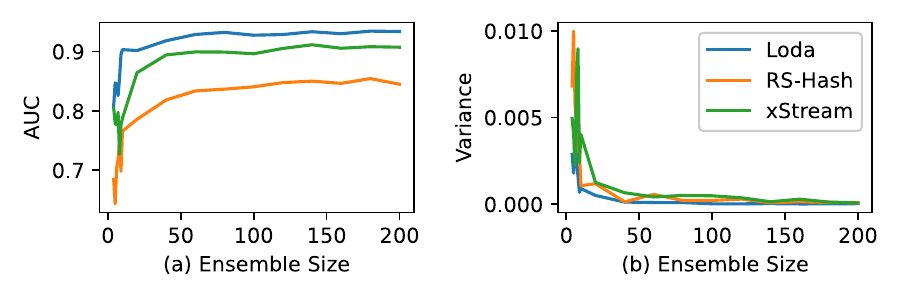}}
\caption{Ensemble Performance Measured on Dataset: Cardio.}
\label{fg:ensemble_cardio}
\end{figure}

Figure~\ref{fg:ensemble_cardio} shows ensemble sizes in the range [3, 200] for Loda, RS-Hash, and xStream respectively. For simplicity, only the Cardio dataset is used to exhibit AUC performance. Figure~\ref{fg:ensemble_cardio} (a) represents the mean AUC and sub-figure (b) shows the variance.

For all detectors, AUC increases with increasing ensemble size before converging to a maximum; the AUC variance shows a decreasing trend before converging to a minimum.
This translates to increasing the number of sub-detectors generally leading to a more reliable result, which follows the underlying principle of ensembles~\cite{Freund1997,Freund1996}.

Furthermore, Table~\ref{tb:model_combo} shows results for different heterogeneous combinations of detectors over our four benchmarks. The rows in this table are divided into two main parts, mean and variance of the AUC, with the columns divided into AUC of Score and Label results. A, B, and C in the second row indicate three detectors (Loda, RS-Hash, and xStream respectively), and the numbers indicate the number of pblocks used. The number of sub-detectors we use for each A, B and C pblock is 35, 25, 20 respectively, details of this value will be described in Section~\ref{se:4_partial_blocks}. For example, A7, B7 and C7 indicate all seven pblocks are utilized for a single type of detector. The fourth one: C223 represents a combination of two pblocks assigned to Loda, two pblocks for RS-Hash, and the last three pblocks being occupied by xStream. We note that this experiment did not cover all possible combinations. However, the goal of this article is to provide a hardware framework that supports arbitrary model combinations; how to get the best combination scheme for a given benchmark is dataset dependent and beyond the scope of this work. The best result for each dataset are in bold.

For the Cardio dataset, Table~\ref{tb:model_combo} shows that Loda always achieves the best Score AUC mean (0.933) and lower variance (0.00004) than RS-Hash, xStream or any listed combination strategy. However, in the label test, all combined labels can get a higher AUC than any single detector, albeit with higher variance. 

In general, it can be seen that Loda, RS-Hash and xStream perform differently for different datasets and there is no single "best" detector, or "ideal" combination. Combined strategies are not always superior to a single detector.
However, a combined detector typically yields more reliable performance. For labels, the combined detector always returns a higher AUC. We believe this is because if any detector indicates an anomaly, the combined result is also an anomaly. This reduces the possibility of missing an anomaly and introducing a higher variance. We believe the strong dataset dependence coupled with the difficulty of finding the best combination of sub-detectors justifies the need to create a runtime reconfigurable framework that can support multiple detectors and multiple instances of each detector.

\begin{table}
\caption{Model Combination Comparison}
\scalebox{0.55}{
\begin{tabular}{|c|c|c|c|c|c|c|c|c|c|c|c|c|c|c|c|c|c|c|c|}
\hline

\multirow{2}{*}{AUC} & \multirow{2}{*}{Model} & \multicolumn{9}{c|}{\textbf{Score}} & \multicolumn{9}{c|}{\textbf{Label}}\\

\cline{3-20}
 & & \textbf{A7} & \textbf{B7} & \textbf{C7} & \textbf{C223} & \textbf{C232} & \textbf{C322} & \textbf{C331} & \textbf{C313} & \textbf{C133} & \textbf{A7} & \textbf{B7} & \textbf{C7} & \textbf{C223} & \textbf{C232} & \textbf{C322} & \textbf{C331} & \textbf{C313} & \textbf{C133}\\

\cline{1-20}
\multirow{4}{*}{Mean} & cardio & \textbf{0.933} & 0.850 & 0.907 & 0.897 & 0.898 & 0.891 & 0.899 & 0.889 & 0.900 & 0.659 & 0.618 & 0.646 & 0.711 & \textbf{0.721} & 0.705 & 0.708 & 0.706 & 0.715\\

 & shuttle & 0.991 & 0.990 & 0.986 & 0.990 & 0.991 & 0.990 & \textbf{0.992} & 0.990 & 0.991 & 0.927 & 0.950 & 0.933 & 0.974 & 0.970 & 0.970 & 0.972 & 0.974 & \textbf{0.976}\\

 & smtp3 & 0.847 & \textbf{0.856} & 0.834 & 0.848 & 0.849 & 0.848 & 0.851 & 0.851 & 0.850 & 0.743 & 0.717 & 0.500 & 0.757 & 0.755 & \textbf{0.770} & 0.767 & 0.765 & 0.737\\

 & http3 & 0.993 & \textbf{0.995} & \textbf{0.995} & \textbf{0.995} & \textbf{0.995} & \textbf{0.995} & \textbf{0.995} & \textbf{0.995} & \textbf{0.995} & 0.595 & 0.510 & 0.512 & 0.595 & \textbf{0.620} & 0.604 & 0.584 & 0.600 & 0.594\\
\cline{1-20}
\multirow{4}{*}{\makecell[c]{Variance\\($\times$ $10^{-3}$)}} & cardio & \textbf{0.04} & 0.2 & 0.05 & 0.07 & 0.06 & 0.09 & 0.06 & 0.23 & 0.05 & 0.13 & 0.14 & 0.09 & \textbf{0.05} & 0.15 & 0.14 & 0.15 & 0.31 & 0.21\\

 & shuttle & 0.005 & \textbf{0.000} & 0.092 & 0.001 & 0.002 & 0.001 & \textbf{0.000} & 0.002 & 0.001 & 0.41 & 0.06 & 3.04 & 0.03 & 0.07 & 0.06 & \textbf{0.01} & 0.03 & 0.02\\

 & smtp3 & \textbf{0.01} & 0.05 & 1.75 & 0.08 & 0.03 & 0.06 & 0.04 & 0.04 & 0.02 & 0.62 & \textbf{0.000} & \textbf{0.000} & 1.96 & 1.95 & 0.93 & 0.89 & 1.03 & 0.71\\

 & http3 & 0.0011 & \textbf{0.0001} & 0.0003 & 0.0003 & \textbf{0.0001} & \textbf{0.0001} & \textbf{0.0001} & \textbf{0.0001} & 0.0002 & 5.01 & 0.13 & \textbf{0.12} & 8.32 & 7.76 & 5.46 & 3.0 & 2.46 & 4.82\\
\hline

\end{tabular}}
\label{tb:model_combo}
\end{table}

\subsection{Pblock Assignment and FPGA Implementation}
\label{se:4_partial_blocks}
Table~\ref{tb:resource_partition} shows a breakdown of resource utilization for the blocks in Figure~\ref{fg:fpga_layout}. RP-1 to RP-7 account for the majority of the FPGA resources and are used for sub-detector implementation. The resource distribution of these blocks is not uniform since the floorplan was manually created to pass the Design Rule Check (DRC) due to the nonuniform resources located on the target FPGA (ZCU111).
The resources for blocks to combine the results, noted as combo-pblocks, are minimal, this being LUT (0.63\%), DSP (0.75\%), BRAM (0.80\%), and FF (0.63\%). Overall, the partial reconfigurable blocks and two static switch blocks utilize 57.73\% LUT, 52.69\% DSP, 55.37\% BRAM, and 57.74\% FF resources. The remaining area is used to implement the remaining interface, including DMAs, AXI-Interconnect, DFX-decoupler etc.  

\begin{table}
  \caption{Resource Partition of FPGA Blocks}
  \scalebox{0.8}{
  \begin{tabular}{lcccc}
    \toprule
    Blocks & LUT & DSP & BRAM & FF\\
    \midrule
    RP-1 & 6.73\% & 4.49\% & 6.67\% & 6.73\%\\
    RP-2 & 8.57\% & 7.54\% & 8.52\% & 8.57\%\\
    RP-3 & 6.24\% & 6.46\% & 6.39\% & 6.24\%\\
    RP-4 & 6.72\% & 4.49\% & 6.67\% & 6.72\%\\
    RP-5 & 6.24\% & 6.46\% & 6.39\% & 6.24\%\\
    RP-6 & 8.74\% & 8.24\% & 8.15\% & 8.74\%\\
    RP-7 & 7.32\% & 7.30\%  & 7.22\% & 7.32\%\\
    RP-combo1 & 0.72\% & 0.56\% & 0.74\% & 0.72\%\\
    RP-combo2 & 0.59\% & 0.84\% & 0.83\% & 0.59\%\\
    RP-combo3 & 0.59\% & 0.84\% & 0.83\% & 0.59\%\\
    Switch-1 & 3.46\% & 4.49\% & 2.96\% & 3.46\%\\
    Switch-2 & 1.81\% & 0.98\% & 0\% & 1.82\%\\
    DMAs & 2.25\% & 0\% & 1.30\% & 0.48\%\\
    DFX-Decouplers & 0.04\% & 0\% & 0\% & 0.008\%\\
    AXI-InterConnect & 0.67\% & 0\% & 0\% & 0.58\%\\
    AXI-SmartConnect & 2.41\% & 0\% & 0\% & 1.61\%\\
    ALL & 62.5\% & 52.69\% & 56.67\% & 60.42\%\\
  \bottomrule
\end{tabular}}
\label{tb:resource_partition}
\end{table}

\begin{table}
  \caption{Resource of 35 Loda, 25 RS-Hash, and 20 xStream for Cardio}
  \scalebox{0.8}{
  \begin{tabular}{lccccc}
    \toprule
    Detector & LUT & DSP & BRAM & FF\\
    \midrule
    Loda-35 & 16783(63.4\%) & 122(44.2\%) & \textbf{54.5(79.0\%)} & 11478(21.7\%) \\
    RS-Hash-25 & \textbf{23732(89.6\%)} & 68(24.6\%) & 50(72.5\%) & 14012(26.5\%)\\
    xStream-20 & \textbf{23908(90.3\%)} & 80(29.0\%) & 60(87.0\%) & 12617(23.8\%)\\
    \textbf{RP-3} & \textbf{26480} & \textbf{276} & \textbf{69} & \textbf{52960} \\
  \bottomrule
\end{tabular}}
\label{tb:resource_max_ad}
\end{table}

An important issue affecting achievable parallelism is how many sub-detectors can be assigned to a pblock. We determine this number by using the smallest pblock (RP-3) as the target and calculate the maximum ensemble size for Loda, RS-Hash and xStream. This exercise lead to ensemble size of $R=35$ for Loda, $R=25$ for RS-Hash, and $R=20$ for xStream in each AD-pblock. The resources required are shown in Table~\ref{tb:resource_max_ad}. Thus if we utilize all seven AD-pblocks to implement a homogeneous type of detector, the current configuration supports a maximum of 245 Loda, 175 RS-Hash, or 140 xStream sub-detectors.

\subsection{Speed, Accuracy and Power Comparison}
\label{se:4_latency_accuracy}
We achieve 80\%-90\% logic use of all seven partial blocks in fSEAD  on the ZCU111 board for homogeneous detect implementations of Loda (245 sub-detectors), RS-Hash (175 sub-detectors) and xStream (140 sub-detectors). This configuration was configured using the PYNQ environment for testing of  fSEAD performance. The FPGA was operated at a clock rate of 188~MHz. 
We also implemented a multi-threaded GCC version of the detectors with the same parameters and scale as fSEAD for comparison.

\begin{table}
  \caption{Accuracy and Execution Time Comparison between fSEAD and CPU for Detector: Loda}
  \scalebox{0.755}{
  \begin{tabular}{lccccrrc}
    \toprule
    Dataset & AUC-S(CPU) & \textbf{AUC-S(FPGA)} & AUC-L(CPU) & \textbf{AUC-L(FPGA)} & Ex Time(CPU) & \textbf{Ex Time(FPGA)} & Speed-up\\
    \midrule
    Cardio & 0.9310 & 0.9311 & 0.6447 & 0.6412 & 13 ms & 4.63 ms & 2.81$\times$\\
    Shuttle & 0.9923 & 0.9914 & 0.9490 & 0.9432 & 147 ms & 34.23 ms & 4.29$\times$\\
    SMTP-3 & 0.8501 & 0.8506 & 0.7666 & 0.7499 & 222 ms & 39.31 ms & 5.65$\times$\\
    HTTP-3 & 0.9937 & 0.9936 & 0.6415 & 0.6336 & 1396 ms & 228.25 ms & 6.12$\times$\\
  \bottomrule
\end{tabular}}
\label{tb:cpu_fpga_loda}
\end{table}

\begin{table}
  \caption{Accuracy and Execution Time Comparison between fSEAD and CPU for Detector: RS-Hash}
  \scalebox{0.755}{
  \begin{tabular}{lccccrrc}
    \toprule
    Dataset & AUC-S(CPU) & \textbf{AUC-S(FPGA)} & AUC-L(CPU) & \textbf{AUC-L(FPGA)} & Ex Time(CPU) & \textbf{Ex Time(FPGA)} & Speed-up\\
    \midrule
    Cardio & 0.8546 & 0.8524 & 0.6310 & 0.6274 & 15 ms & 4.87 ms & 3.08$\times$\\
    Shuttle & 0.9915 & 0.9910 & 0.9543 & 0.9560 & 168 ms & 35.80 ms & 4.69$\times$\\
    SMTP-3 & 0.8525 & 0.8513 & 0.7166 & 0.7166 & 260 ms & 39.63 ms & 6.56$\times$\\
    HTTP-3 & 0.9944 & 0.9944 & 0.5065 & 0.5067 & 1490 ms & 228.29 ms & 6.53$\times$\\
  \bottomrule
\end{tabular}}
\label{tb:cpu_fpga_rshash}
\end{table}

\begin{table}
  \caption{Accuracy and Execution Time Comparison between fSEAD and CPU for Detector: xStream}
  \scalebox{0.755}{
  \begin{tabular}{lccccrrc}
    \toprule
    Dataset & AUC-S(CPU) & \textbf{AUC-S(FPGA)} & AUC-L(CPU) & \textbf{AUC-L(FPGA)} & Ex Time(CPU) & \textbf{Ex Time(FPGA)} & Speed-up\\
    \midrule
    Cardio & 0.9229 & 0.9222 & 0.6467 & 0.6435 & 18 ms & 4.82 ms & 3.73$\times$\\
    Shuttle & 0.9914 & 0.9905 & 0.9688 & 0.9680 & 250 ms & 40.62 ms & 6.15$\times$\\
    SMTP-3 & 0.8077 & 0.8076 & 0.7167 & 0.7167 & 366 ms & 50.99 ms & 7.18$\times$\\
    HTTP-3 & 0.9947 & 0.9948 & 0.5067 & 0.5069 & 2460 ms & 297.85 ms & 8.26$\times$\\
  \bottomrule
\end{tabular}}
\label{tb:cpu_fpga_xstream}
\end{table}

The detector system in fSEAD is configured with the topology shown in Figure ~\ref{fg:combination_scheme}(c). An averaging-based  combination and an OR-Gate-based label combination techniques were used. The ap\_fixed<32,16,AP\_TRN,AP\_WRAP> type available in Xilinx Vivado HLS~\cite{vivado_hls} was used for all inner non-integer operations. This has a word-length of 32-bit with 16 bits representing the integer part and 16 bits representing the fraction. Convergent rounding (AP\_TRN) and saturating arithmetic (AP\_WRAP) were used. In order to facilitate the use of the float32 type in the NUMPY library~\cite{numpy} for streaming data transfer with each module on the FPGA, all fSEAD IP interfaces are converted to float32. This does not cause a significantly increase in resource consumption.

\begin{figure}
\centerline{\includegraphics[width=0.4\linewidth]{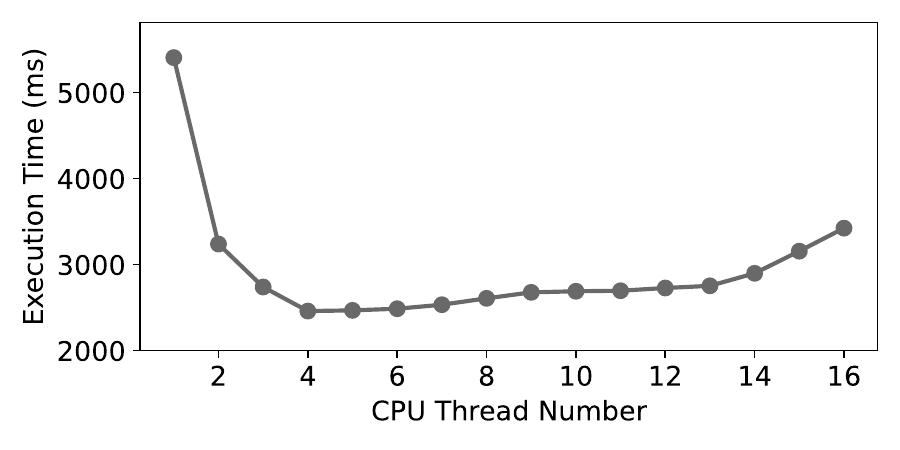}}
\caption{Multi-threaded CPU Implementation for xStream for HTTP-3}
\label{fg:multi_thread}
\end{figure}

Tables~\ref{tb:cpu_fpga_loda} through~\ref{tb:cpu_fpga_xstream} show the performance comparison of Loda, RS-Hash and xStream detectors in terms of AUC and execution time on the ZCU111 board and CPU, respectively. The execution time test uses the \textit{time()} function in Python to measure the time from the start of the input DMA transfer to when all data is obtained from the output DMA. The GCC implementation was executed on a multi-threaded CPU on the target PC, and the \textit{time} command in $bash$ used to measure execution time. The pthread library is used to support the multi-threaded C implementation. Considering that each sub-detector in the ensemble is data-independent, we equally distribute the same number of sub-detectors to each CPU thread. The sub-score computed by multiple threads requires a synchronization operation at the end of each sample to compute the average score of the ensemble. The $pthread\_mutex\_lock$ and $pthread\_mutex\_unlock$ functions are placed between different threads to guarantee the streaming mode execution. The target CPU has 8 cores and 16 hardware threads; we conduct a test of switching the thread number from 1 to 16. Figure~\ref{fg:multi_thread} shows the speedup results with different numbers of threads on the longest test-case: xStream for HTTP-3, 4-thread always get the best speedup. We believe this is due to synchronization overheads introduced by the $mutex$ scheme limiting performance when the number of threads is greater than 4 ($mutex$ is called in every streaming execution). Based on this result, all the following CPU experiments results are measured from 4-thread configuration.

\begin{figure}
\centerline{\includegraphics[width=1.0\linewidth]{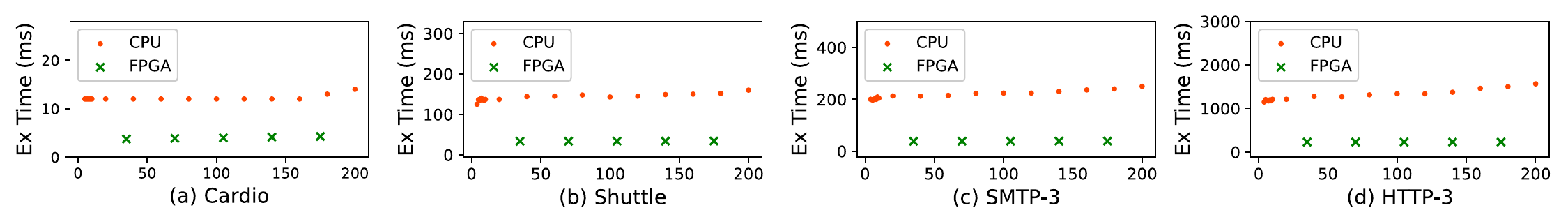}}
\caption{Execution Time Comparison of fSEAD and CPU for Detector: Loda.}
\label{fg:execttion_time_loda}
\end{figure}

\begin{figure}
\centerline{\includegraphics[width=1.0\linewidth]{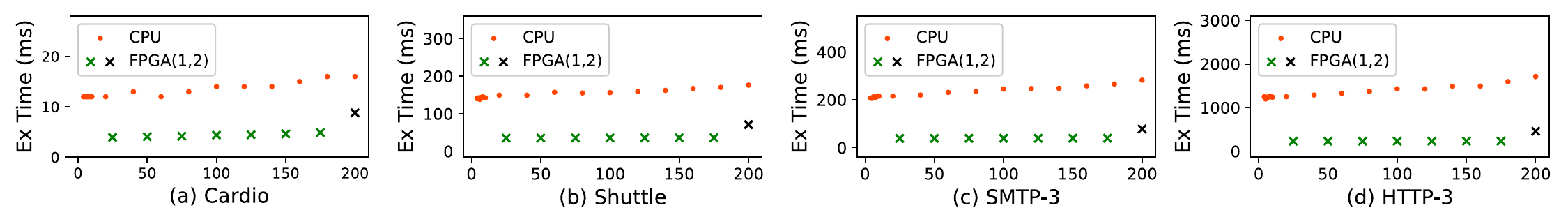}}
\caption{Execution Time Comparison of fSEAD and CPU for Detector: RS-Hash.}
\label{fg:execttion_time_rshash}
\end{figure}

\begin{figure}
\centerline{\includegraphics[width=1.0\linewidth]{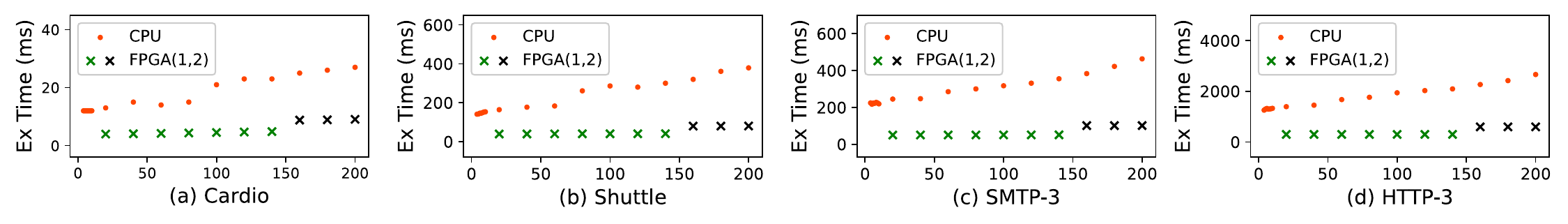}}
\caption{Execution Time Comparison of fSEAD and CPU for Detector: xStream.}
\label{fg:execttion_time_xstream}
\end{figure}

In terms of accuracy, very similar scores were obtained on CPU and FPGA platforms. 
This indicates the ap\_fixed<32,16> variable type used can provide sufficient accuracy as the float32 variable type of CPU implementation. The minor differences are due to the accumulation of errors during the cumulative calculation of the accuracy differences generated by each ap\_fixed and float32 type sub-detector. Future developers may wish to explore minimising resource consumption further by reducing precision.

On the smallest dataset, Cardio, it can be observed that the FPGA obtains a speed-up of 2.81 for Loda, 3.08 for RS-Hash, and 3.73 times for xStream. The speedup ratio increases as the size and dimension of the dataset increases. On the largest dataset, HTTP-3, the three detectors achieve speedups of 6.12, 6.53 and 8.26 respectively. The highest speedup 8.26 is achieved by xStream for dataset: HTTP-3. For smaller datasets, the transfer time from the Linux OS-based host ARM processor to the FPGA becomes the bottleneck. As fSEAD is optimised for large datasets, this is not a significant issue.

It is worth mentioning that since the ensemble on fSEAD is based on sub-detector-level parallelism, its latency is not significantly affected by the ensemble size in the case that FPGA resources are sufficient to implement the required ensemble in parallel. In contrast, the ensemble implementation on GCC uses a \textit{for} loop to iterate the computation of each sub-detector, so its execution time increases proportionally with the increase of the ensemble size. Figures~\ref{fg:execttion_time_loda} to Figures~\ref{fg:execttion_time_xstream} show the execution times of the three detectors on the CPU and FPGA for multiple sets of experiments and different ensemble sizes. The red dots indicating CPU execution time show a linear increase with  ensemble size. The green crosses indicate the execution time on FPGA, while the black cross is the results of two FPGA executions.
In all cases, significant speed-ups are achieved over the CPU and limited only by FPGA resources.

\begin{table}
\caption{Operation Number of fSEAD}
\scalebox{0.8}{
\begin{threeparttable}[t]

\begin{tabular}{|c|c|}
\hline
Detectors & Operation Number \\
\hline
Loda & $OP = N*(2Rd + 7R + 2)$ \\
\hline
RS-Hash & $OP = N*(5Rdw + 4Rd + 11Rw + R + 2)$ \\
\hline
xStream & $OP = N*(2Rdk + 5Rdw + 15Rw + 2R + 2)$ \\
\hline

\end{tabular}
\begin{tablenotes}
    \footnotesize
    \item[1] \(N\): the length of the input dataset.
    \item[2] \(d\): the dimension of input dataset.
    \item[3] \(R\): the ensemble size.
    \item[4] \(w\): the hash functions number in CMS.
    \item[5] \(k\): the projection size of xStream.
\end{tablenotes}

\end{threeparttable}}
\label{tb:OP}
\end{table}

\begin{table}[]
\caption{GOPS Comparison of CPU and fSEAD}
\scalebox{0.8}{
\begin{tabular}{|c|c|c|c|c|c|c|c|c|c|}
\hline

\multirow{2}{*}{GOPS} & \multicolumn{4}{c|}{\textbf{CPU}} & \multicolumn{4}{c|}{\textbf{fSEAD}}\\

\cline{2-9}
 &\textbf{Cardio} & \textbf{Shuttle} & \textbf{SMTP-3} & \textbf{HTTP-3} & \textbf{Cardio} & \textbf{Shuttle} & \textbf{SMTP-3} & \textbf{HTTP-3}\\
\hline
 Loda & 1.690 & 2.049 & 1.402 & 0.776 & 4.748 & 8.789 & 7.924 & 4.748\\
\hline
 RS-Hash & 6.772 & 6.353 & 4.197 & 4.331 & 20.858 & 29.797 & 27.533 & 28.282\\
\hline
 xStream & 15.427 & 11.050 & 6.623 & 5.878 & 57.544 & 67.959 & 47.554 & 48.551\\
\hline

\end{tabular}}
\label{tb:GOPS}
\end{table}

\begin{figure}
	\centering
	\begin{minipage}{0.49\linewidth}
		\centering
		\includegraphics[width=0.9\linewidth]{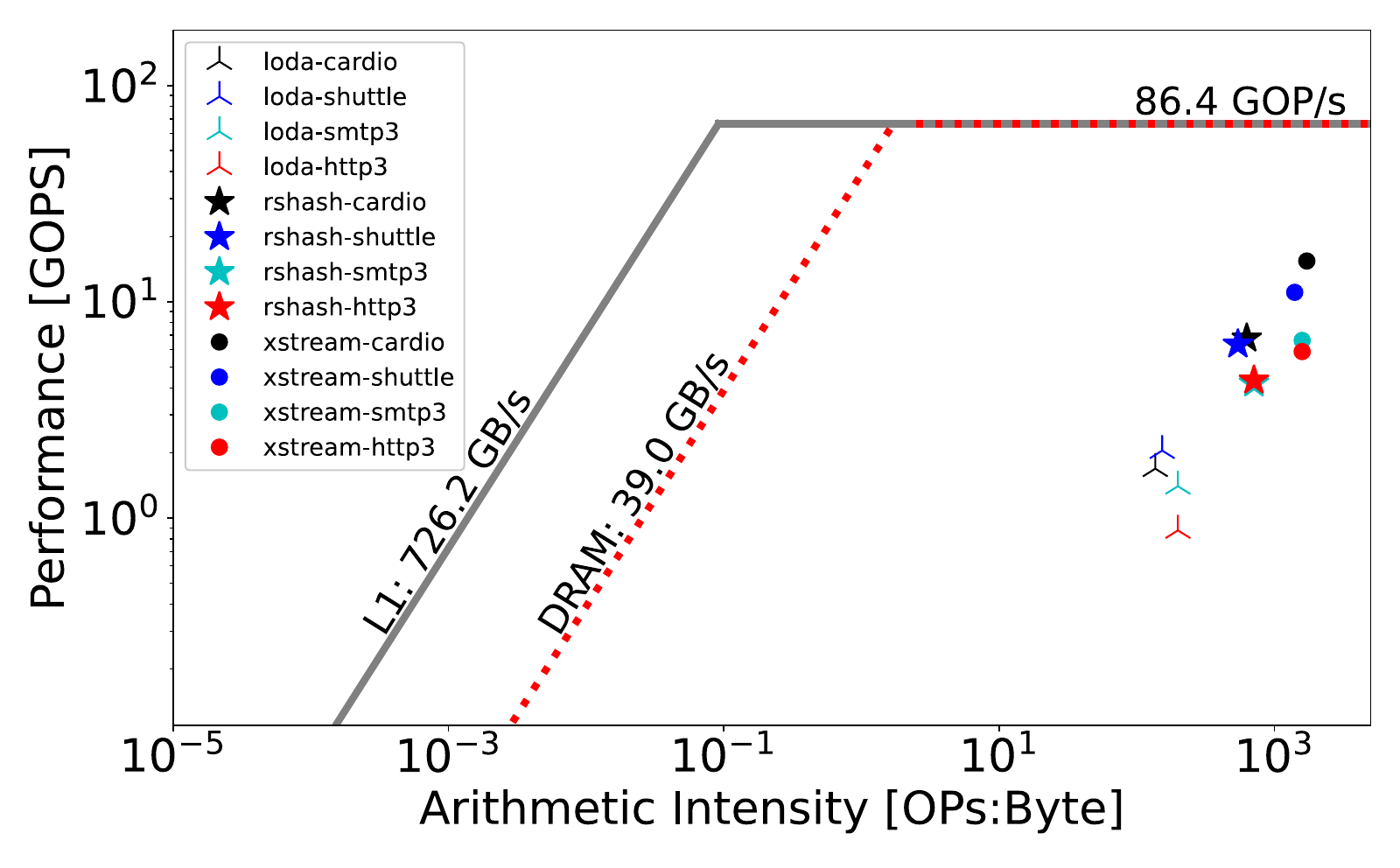}
		\caption{Roofline Model on CPU.}
		\label{fg:roofline_cpu}
	\end{minipage}
	\begin{minipage}{0.49\linewidth}
		\centering
		\includegraphics[width=0.9\linewidth]{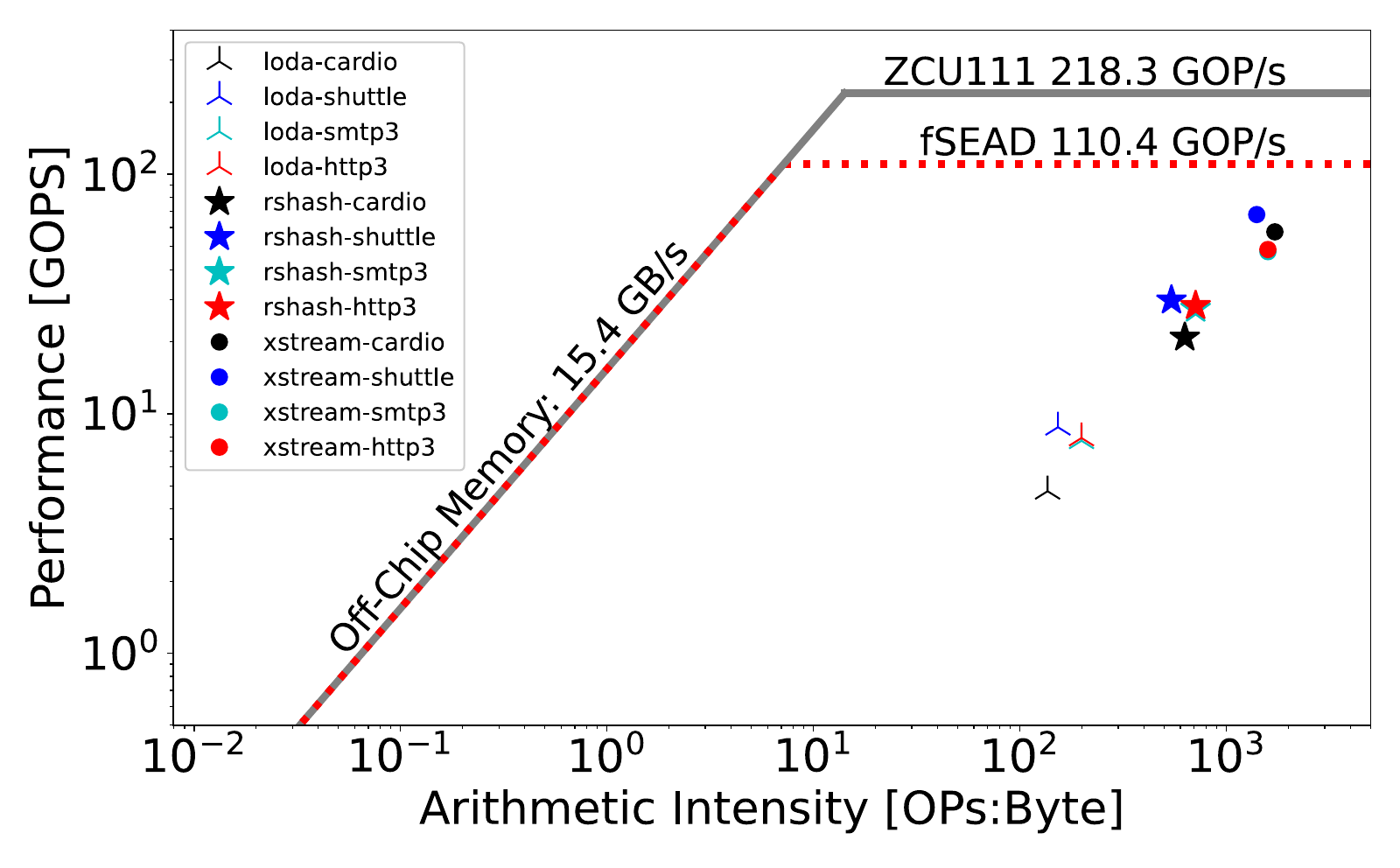}
		\caption{Roofline Model on FPGA.}
		\label{fg:roofline_fpga}
	\end{minipage}
\end{figure}

We also use the roofline model to estimate the degree to which our target anomaly detector methods have been optimised. This describes an application's achieved performance and arithmetic intensity against the machine's maximum achievable performance in terms of memory bandwidth and peak computational performance. The CPU roofline model is measured on the Intel Advisor 2022~\cite{intel_advisor}. Billions of operations per second (GOPS) is used as the metric for Performance (y axis in Figure~\ref{fg:roofline_cpu}) and operations (OPs) per byte is used for Arithmetic intensity (x axis in Figure~\ref{fg:roofline_cpu}). 
The L1/DRAM bandwidth roofline represents the maximum amount of bytes that can get read or written for a given arithmetic intensity.
For FPGA roofline models, we use the method in Reference~\cite{semiroofline} to calculate the roofline chart in Figure~\ref{fg:roofline_fpga}. The arithmetic intensity (x axis) is the number of arithmetic operations performed for each byte read or written to off-chip memory (we assume 13.4 GB/s off-chip memory bandwidth~\cite{xilinx_zcu111}). The performance (y axis) is calculated in GOPS. 

For the three detectors, Table~\ref{tb:OP} shows the expressions we used to estimate the total number of  operations for the execution of a target dataset. Using this and the execution time from Table~\ref{tb:cpu_fpga_loda} to Table~\ref{tb:cpu_fpga_xstream}, we compute the GOPS in Table~\ref{tb:GOPS}. We use the highest GOPS (67.959 GOPS of xStream for Shuttle) to estimate the compute-bound performance for the target FPGA (ZCU111) and fSEAD structure respectively. The xStream for Shuttle occupies 132391 LUTs, 476 DSPs and 79485 FFs which takes up 31.13\% of the total FPGA resources and 61.57\% of the fSEAD partial blocks resources. From this, we calculate the compute-bound performance for FPGA and fSEAD as 218.3 GOP/s and 110.4 GOP/s respectively. 

While no algorithms reach the boundary of the roofline, xStream is closer to the computational boundary than Loda and RS-Hash, as seen in Figure~\ref{fg:roofline_cpu} and Figure~\ref{fg:roofline_fpga}. We believe this is a function of the algorithms: the current three detectors are not extremely computationally intensive, but xStream has more vector and matrix multiplication operations among the current AD library.

\begin{figure}
\centerline{\includegraphics[width=0.4\linewidth]{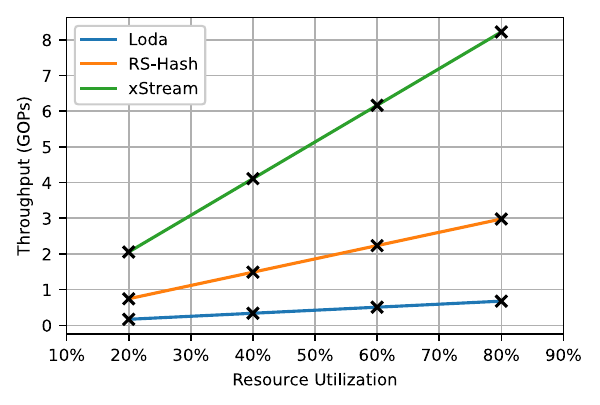}}
\caption{Example of the Scalability inside Single Partial Block: RP-1}
\label{fg:throughput_rp1}
\end{figure}

Figure~\ref{fg:throughput_rp1} shows scalability of a single pblock, RP-1, using the Cardio dataset with 20\% to 80\% resource utilization for Loda, RS-Hash and xStream. Working at the same 188~MHz clock frequency, the throughput of three detectors can be seen to vary linearly with the resource utilization of RP-1. This linear scalability enables one to quickly identify potential ensembles that will fit on a given FPGA and estimate their throughput. One could then perform software experiments, such as those in Table~\ref{tb:model_combo}, to identify which of these potential ensembles provides the best accuracy for the target dataset.

\begin{figure}
	\centering
	\begin{minipage}{0.49\linewidth}
		\centering
		\includegraphics[width=0.78\linewidth]{./figures_tables/power_chip.pdf}
		\caption{Chip Power Consumption.}
		\label{fg:power_chip}
	\end{minipage}
	\begin{minipage}{0.50\linewidth}
		\centering
		\includegraphics[width=1.0\linewidth]{./figures_tables/power_system.pdf}
		\caption{System Power Test-bed.}
		\label{fg:power_system}
	\end{minipage}
\end{figure}

The power consumption of fSEAD is separated into chip power and system power, Figure~\ref{fg:power_chip} and Figure~\ref{fg:power_system} show the measured chip and system power consumption from Xilinx Vivado Tool~\cite{vivado_power_analysis} and EcoFlow RIVER Max Portable Power Station~\cite{ecoflow} respectively. In Figure~\ref{fg:power_system}, the idle system power measured on EcoFlow is 30 Watts and the working system power (35 Watts) is gained by configuring the PYNQ to invoke all pblocks for xStream with the biggest dataset HTTP-3, which matches the dynamic power of 5.232 Watts, as demonstrated in Figure~\ref{fg:power_chip}. Intel(R) Core(TM) i7-10700F CPU power is measured with the powerstat command using Running Average Power Limit (RAPL) domains. 120 samples of power measurements with 0.5 second step are averaged over the duration of one minute time. The CPU idle power is 7.90 Watts and the CPU working power for xStream with the biggest dataset HTTP-3 is 51.23 Watts. The dynamic CPU power (43.33 Watts) is more than $8\times$ higher than the fSEAD dynamic power consumption on ZCU111 FPGA (5.232 Watts).

\subsection{Partial Reconfiguration}
\label{se:4_dfx_speed_generic_useage}
Although the reconfiguration of each partial region in fSEAD can often be done when the system is idle, we measure the overhead of partial reconfiguration in the PYNQ framework. The horizontal axis of Table~\ref{tb:DFX_Ex Time} shows all pblocks and the vertical axis shows the direction of reconfiguration for bitstream downloads. For example, $Function \rightarrow Identity$ indicates that the configured logic is $Function$ which is overwritten with $Identity$. We have chosen a common function module: $Loda\_Cardio$ and $Averaging$ for (RP-1 to RP-7) and (COMBO1 to COMBO3) respectively as the $Function$ module. $Identity$ is a design where the input is simply passed to the output. The objective is to evaluate the impact of the hardware logic size of the target bitstream on the reconfiguration time cost. Referring to the size of each pblock provided in Table~\ref{tb:resource_partition} (all seven AD-pblock resources are larger than those of COMBO1 to COMBO3; among the seven AD-pblocks, RP-3 has the least programmable resources and RP-6 the most; while COMBO1 occupies the largest area of the three COMBO blocks). 

The latencies reported in Table~\ref{tb:DFX_Ex Time} are the total partial reconfiguration latency. This will be a function of the size and location of the region, as well as the size of the partial bitstream, and the chosen FPGA. Since we do not have the ability to modify the chosen FPGA, in this section, we explore how these other factors impact reconfiguration time. The results in Table~\ref{tb:DFX_Ex Time} show that the pblock with larger area takes a little more time to reconfigure, e.g. RP-6 takes 609.6 ms for updating $Loda\_Cardio$ with $Identity$. The pblock with the smallest area uses the shortest time i.e. COMBO3 used 579.8 ms to reconfigure $Loda\_Cardio$ to $Identity$. In addition, the complexity of the target bitstream has a slight impact on the reconfiguration time. Apart from RP-2, RP-3 and COMBO3, a general trend indicates that the simpler the logic of the target bitstream (in this case, $Identity$), the lower the reconfiguration time.

\begin{table}
  \caption{Partial Reconfiguration Time Cost (Unit is ms) of the Different Pblocks}
  \scalebox{0.82}{
  \begin{tabular}{lcccccccccc}
    \toprule
    BIT download directions & RP-1 & RP-2 & RP-3 & RP-4 & RP-5 & RP-6 & RP-7 & COMBO1 & COMBO2 & COMBO3\\
    \midrule
    $Function\rightarrow Identity$ & 607.8 & 606.1 & 604.5 & 606.1 & 608.9 & \textbf{609.6} & 609.5 & 587.2 & 582.7 & \textbf{579.8}\\
    $Identity\rightarrow Function$ & 606.3 & 611.3 & 607.2 & 606.0 & 606.9 & 608.1 & 607.5 & 582.9 & 580.1 & 581.9\\
    
  \bottomrule
\end{tabular}}
\label{tb:DFX_Ex Time}
\end{table}

\begin{figure}
\centerline{\includegraphics[width=0.7\linewidth]{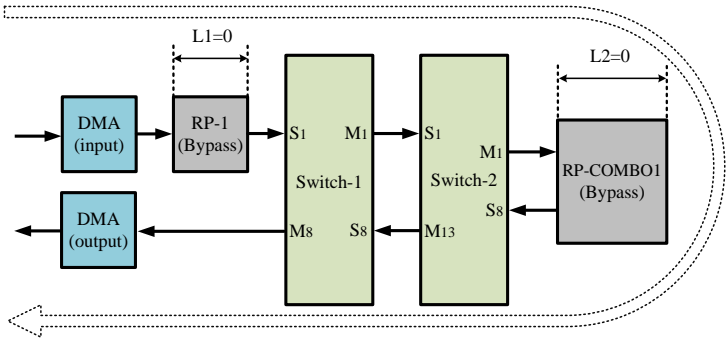}}
\caption{Example of fSEAD Channel with Empty Logic.}
\label{fg:empty_loopback}
\end{figure}

While fSEAD is focused on ensemble-centric anomaly detection, different real-life detectors in the literature can be incorporated into this library for higher flexibility and salable combinations. Moreover, it has the potential to be used for more general machine learning applications by adding different modules. To evaluate the default latency of fSEAD interconnections, since this is a key metric that could affect design decisions, Figure~\ref{fg:empty_loopback} illustrates a data-path where each pblock simply copies its input to output (the "Bypass" in Figure~\ref{fg:empty_loopback} is the same as $Identity$ in Table~\ref{tb:DFX_Ex Time}). The execution time, which is a measure of system latency overhead, is \textbf{0.80 ms}. Thus for pblocks with latency L1 on the left and L2 on the right, the maximum latency of the system is \textbf{$\approx$ 0.80+L1+L2 (ms)}. We also measure the shorter DMA latency where the data-path include: DMA (input), L1, Switch-1 and DMA (output), the latency is \textbf{0.77 ms}. This reflects that the default system latency is dominated by the Linux OS-based PYNQ framework, rather than by the routing latency of switches.

\section{Conclusions and Future work}
\label{se:5_conclusion}
In this article, we proposed a flexible computing architecture consisting of multiple partially reconfigurable regions, called pblocks which are interconnected via the AXI-Streaming Switches. The scheme enables parallel constructions of repeating blocks to be easily scaled to fill the FPGA. We also demonstrated a concrete application of this architecture to streaming anomaly detection using ensembles (fSEAD). fSEAD allows complex and more powerful anomaly detectors to be composed from simple pblocks in an arbitrary fashion. 
Careful floorplanning of the pblocks and switches, together with bus-level pipelining minimise routing delay and allow timing closure to be achieved. Through a number of experiments involving ensembles of three sub-detectors (Loda, RS-Hash and xStream), we demonstrate 3 to 8$\times$ speedup compared with a CPU.

Future work will extend this architecture in several directions. First, continuing with the anomaly detector application, we will create module generators which can automatically generate optimized HLS directives and detector parameters such as ensemble size. We will also support more anomaly detection methods to enrich the existing library. Finally, we plan to use the  proposed composable architecture for a wider range of applications, e.g. classification and regression tasks. 

\begin{acks}
Binglei Lou gratefully acknowledges financial support from the China Scholarship Council.
\end{acks}

\bibliographystyle{ACM-Reference-Format}
\bibliography{ref}

\appendix

\end{document}